\documentclass{jaa}

\usepackage{hyperref}
\usepackage{times,graphicx,amssymb, amsmath}
\usepackage{epsfig}

\begin{document}\sloppy

\title{Nonlinear power spectrum in clustering and smooth dark energy models beyond the BAO scale}


\author{Bikash R. Dinda\textsuperscript{1,*}}
\affilOne{\textsuperscript{1}Centre for Theoretical Physics, Jamia Millia Islamia, New Delhi-110025, India.}


\twocolumn[{

\maketitle

\corres{bikashd18@gmail.com}

\msinfo{---}{---}

\begin{abstract}
We study the nonlinear effects of the clustering and smooth quintessence. We present numerical and also approximate semi-analytical expressions of nonlinear power spectrum both for clustering and smooth dark energy models beyond the Baryon Acoustic Oscillations (BAO) scale. This approximation is motivated by the extension of the resummation method of Anselmi $\&$ Pietroni (J Cosmol Astro-Part Phys 12:13, 2012.
\url{arXiv:1205.2235}) for the dark energy models with evolving equation of state. The results of this scheme allow us for the prediction of the nonlinear power spectrum in the mildly nonlinear regime up to few percentage accuracies compared to the other available tools to compute the nonlinear power spectrum for the evolving dark energy models.
\end{abstract}


}]


\doinum{12.3456/s78910-011-012-3}
\artcitid{\#\#\#\#}
\volnum{000}
\year{0000}
\pgrange{1--}
\setcounter{page}{1}
\lp{1}

\section{Introduction}

Many cosmological observations like Supernova Type-Ia observations (A. G. Riess 1998, S. Perlmutter 1999), cosmic microwave background observations (D. N. Spergel 2003, G. Hinshaw 2003), baryon acoustic oscillations measurements (T. Delubac 2015,  M. Ata 2017) strongly support the late time cosmic acceleration. Theoretically, to explain the late time cosmic acceleration, either we have to consider some exotic matter called dark energy (I.  Zlatev 1999, P. J. Steinhardt 1999, Caldwell R. R. 2005, Eric V. Linder 2006, Shinji Tsujikawa 2010, Scherrer R. J. 2008, Bikash R. Dinda 2016, T. Chiba 2009) or we have to modify the general theory relativity (T. Clifton 2012, K. Hinterbichler 2012, C. de Rham 2012, C. de Rham 2014, A. De Felice 2010). The simplest dark energy model is the $\Lambda$CDM model (Ade P. A. R. 2016a). In $\Lambda$CDM model, the equation of state of the dark energy is constant and it is $-1$. However, in literature, there are many dark energy models which support cosmological observations at some significant confidence level (Edmund J. Copeland 2006). One such model is the quintessence model (I.  Zlatev 1999, P. J. Steinhardt 1999, Caldwell R. R. 2005, Eric V. Linder 2006, Shinji Tsujikawa 2010, Scherrer R. J. 2008, Bikash R. Dinda 2016, T. Chiba 2009).

In general, quintessence can cluster only on super-horizon scales. But in literature, sometimes clustering quintessence are considered (P. Creminelli 2009, P. Creminelli 2010). In this case quintessence has vanishing speed of sound (P. Creminelli 2009, P. Creminelli 2010, P. Creminelli 2006, L. Senatore 2005, C. Cheung 2008, S. DeDeo 2003, J. Weller 2003). Since the clustering quintessence can cluster on all scales, it adds the extra effect on the observables like power spectrum both on large and small scales. To study the signatures of the clustering quintessence on small scales, modeling of the nonlinear structure formation is extremely important for the clustering dark energy models.

Among different techniques to study the nonlinear power spectrum (Zhaoming Ma 2007, Takahashi 2012, M. Crocce 2006a, S.~Matarrese 2007, S.~Anselmi 2011b, M. Crocce 2006b, G.D’Amico 2011, P. Creminelli 2010), in this paper, we focus on the resummation technique (S. Anselmi 2012, S. Anselmi 2014). The resummation method was first introduced by Anselmi and Pietroni in 2012 to compute the nonlinear power spectrum beyond the BAO scale for the $\Lambda$CDM model of dark energy (S. Anselmi 2012). Their results agree with the N-body simulations at the few percent levels up to $k \lesssim 1 h Mpc^{-1}$ at $z \gtrsim 0.5$ and up to $k \lesssim 0.6 h Mpc^{-1}$ at $z < 0.5$ respectively (S. Anselmi 2012).

In (S. Anselmi 2014), the authors extended the resummation scheme for the other dark energy models with the constant equation of state (both for smooth and clustering quintessence). The results agree at few percent levels up to $k \lesssim 0.6 h Mpc^{-1}$ (S. Anselmi 2014).

However, the natural choice of dark energy models (especially quintessence models)  possesses evolving equation of state of dark energy. In this paper, we extend the resummation scheme for the dark energy models where the equation of state of the dark energy is evolving. We have considered popular CPL (Chevallier-Polarski-Linder) parametrization (M. Chevallier 2001, E. V. Linder 2003). One can apply this method to any evolving dark energy model.

The paper is organized as follows: in section 2 we study the background evolution of the dark energy model (particularly the CPL parametrization); in section 3 we calculate the perturbation equations both for clustering and smooth dark energy; in section 4  we solve these perturbation equations on linear scales; in section 5 we extend the resummation scheme to compute the nonlinear power spectrum; in section 6 we present our results; and finally in section 7 we conclude our work.

\section{Background evolution}

We consider the flat Friedmann-Robertson-Walker (FRW) Universe with two components, total matter (including cold dark matter (CDM)) and dark energy. As we are interested in the late time evolution of the Universe, we can safely ignore the contribution of radiation. For dark energy, we assume the most studied CPL parametrization (M. Chevallier 2001, E. V. Linder 2003) which is described by the equation of state as

\begin{equation}
w = w_{0} + w_{a}(1-a),
\label{eq:eos}
\end{equation}

\noindent
where, $ w_{0} $ and $ w_{a} $ being two model parameters and $ a $ being the scale factor. $ w_{0} $ corresponds to the present day equation of the state of dark energy. $ w_{a} $ describes how the equation of state parameter changes with time from the present day value.

\begin{figure}
\begin{center}
\includegraphics[scale=0.4]{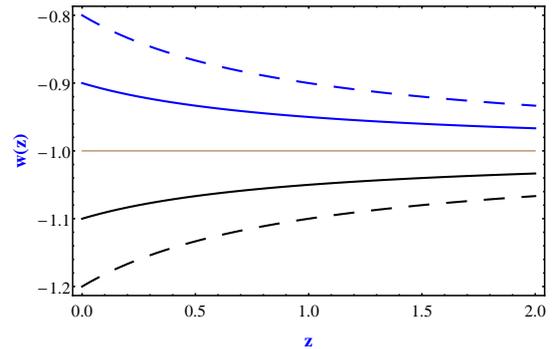}
\end{center}
\caption{\label{fig:wofz} $ w(z) $ vs. $ z $ graphs for the models (1) $w_{0}=-1.2$, $w_{a}=0.2$ (Black-Dashed), (2) $w_{0}=-1.1$, $w_{a}=0.1$ (Black), (3) $\Lambda$CDM (Brown), (4) $w_{0}=-0.9$, $w_{a}=-0.1$ (Blue), and (5) $w_{0}=-0.8$, $w_{a}=-0.2$ respectively.}
\end{figure}

In Fig.~\ref{fig:wofz}, we have plotted $ w(z) $ vs. $ z $ graphs to show how the equation of state of the dark energy evolves with time. We have taken $ 5 $ models throughout in this paper. The models are: (1) $w_{0}=-1.2$, $w_{a}=0.2$ (Black-Dashed), (2) $w_{0}=-1.1$, $w_{a}=0.1$ (Black), (3) $\Lambda$CDM (Brown), (4) $w_{0}=-0.9$, $w_{a}=-0.1$ (Blue), and (5) $w_{0}=-0.8$, $w_{a}=-0.2$ (Blue-Dashed). We use the same colour code throughout (except Fig.~\ref{fig:NBDcmp}). We can see that the first two models have phantom ($w<-1$) behaviour whereas last two models have non-phantom ($w>-1$) behaviour. To mention, 1st model is more phantom than 2nd model and 5th model is more non-phantom than 4th model respectively. All the model parameters are chosen in such a way that at sufficient early time all the models posseses $ w \sim -1 $. This choice has been considered because we want to normalize the perturbation quantities (like growth function, power spectrum etc.) at early matter dominated era (at redshift $ z=z_{in}=1000 $). So, it is better to consider models with same initial equation of state ($w(z_{in})$) to be consistent.

Note that background evolution is the same for both the clustering and smooth dark energy. The difference will come from the perturbation evolution. We discuss this in the next section.

\section{Perturbation equations}

Throughout this paper, we are considering the evolution of the perturbations in the sub Horizon scale where Newtonian perturbation theory is valid. We consider both the clustering and smooth dark energy models described below.

\subsection{Clustering dark energy}

\noindent
First, we consider clustering dark energy with $c_{s}^2 = 0$ which is same as that of the matter component. Hence we can write the following equation governing the evolution of density constrast for matter and dark energy as well as for the velocity perturbation (note that due to idenitical sound speed, both the matter and dark energy have the same peculiar velocity, $ v = v_{m} = v_{Q} $) (E. Sefusatti 2011, G.D’Amico 2011):

\begin{equation}
\dfrac{\partial \delta_{m}}{\partial \tau}+\vec{\nabla}.[(1+\delta_{m})\vec{v}]=0,
\label{eq:CntnMatter}
\end{equation}

\begin{equation}
\dfrac{\partial \delta_{Q}}{\partial \tau}-3 \omega \mathcal{H} \delta_{Q}+\vec{\nabla}.[(1+\omega+\delta_{Q})\vec{v}]=0,
\label{eq:CntnDE}
\end{equation}

\begin{equation}
\dfrac{\partial \vec{v}}{\partial \tau}+ \mathcal{H} \vec{v} + (\vec{v}.\vec{\nabla})\vec{v}=-\vec{\nabla} \Phi,
\label{eq:EulerSame}
\end{equation}

\noindent
and

\begin{equation}
\nabla^{2} \Phi = 4\pi G a^{2} (\delta \rho_{m}+\delta \rho_{Q}) = \dfrac{3}{2} \mathcal{H}^{2} \Omega_{m} (\delta_{m} + \dfrac{\Omega_{Q}}{\Omega_{m}} \delta_{Q}),
\label{eq:Poisson}
\end{equation}

\noindent
where $ \delta \rho_{m} $ $ \& $ $ \delta \rho_{Q} $ and $ \delta_{m} $ $ \& $ $ \delta_{Q} $ are density fluctuations $ \& $ density contrasts for CDM and clustering dark energy respectively, $ \vec{v} $ is the common peculiar velocity field, $ \tau $ is the conformal time, $ \mathcal{H}$ is the conformal Hubble parameter, and $ \Phi $ is the gravitational potential. The matter and dark energy density parameters are given by $\Omega_{m} = \Omega_{m}^{(0)} a^{-3} \left(\frac{H}{H_{0}}\right)^{-2}$ and $\Omega_{Q} = 1 - \Omega_{m}$ respectively. The Hubble parameter is given by $H^2 = H_{0}^2 \left[ \Omega_{m}^{(0)} a^{-3} + (1-\Omega_{m}^{(0)})a^{-3(1+w_{0}+w_{a})} e^{3 w_{a}(a-1)} \right]$, where $ \Omega_{m}^{(0)} $ and $ H_{0} $ are the present day matter density parameter and Hubble parameter respectively. Throughout in this paper we consider $ \Omega_{m}^{(0)} h^{2} = 0.13263 $ and $ H_{0} = 100 h km/s/Mpc $ with $h = 0.72$. Note that the above set of equations are valid for the sub-Hubble limit, which is a good assumption to study dynamics of structure formation.

\noindent
We can further assume velocity fields are irrotational and it can be completely described by its divergence $ \theta=\vec{\nabla} . \vec{v} $ (F. Bernardeau 2002). Taking divergence of Eq. \eqref{eq:EulerSame}, we can write the equation for $ \theta $ as  

\begin{equation}
\dfrac{\partial \theta}{\partial \tau} + \mathcal{H} \theta +\dfrac{3}{2} \mathcal{H}^{2} \Omega_{m} (\delta_{m} + \dfrac{\Omega_{Q}}{\Omega_{m}} \delta_{Q})= - \vec{\nabla}.[(\vec{v}.\vec{\nabla})\vec{v}].
\label{eq:Euler2}
\end{equation}

\noindent
The total density contrast can now be defined as

\begin{equation}
\delta = \dfrac{\delta \rho}{\bar{\rho_{m}}} = \delta_{m} + \dfrac{\Omega_{Q}}{\Omega_{m}} \delta_{Q}.
\label{eq:deltaTotal}
\end{equation}

\noindent
The Poisson equation Eq. \eqref{eq:Poisson} can now be written in terms of total density contrast as

\begin{equation}
\nabla^{2} \Phi = \dfrac{3}{2} \mathcal{H}^{2} \Omega_{m} \delta.
\label{eq:Poisson2}
\end{equation}

\noindent
Using Eqs. \eqref{eq:deltaTotal} and \eqref{eq:Poisson2} we can write down two continuity Eqs. \eqref{eq:CntnMatter} and \eqref{eq:CntnDE} together into a single equation of $ \delta $ and $ \theta $ as

\begin{equation}
\dfrac{\partial \delta}{\partial \tau} +C(\tau) \theta = - \vec{\nabla}.(\delta \vec{v}),
\label{eq:CntnTotal}
\end{equation}

\noindent
where we have introduced a new function $ C(\tau) $ which is given by

\begin{equation}
C(\tau) = 1+(1+\omega)\dfrac{\Omega_{Q}}{\Omega_{m}}.
\label{eq:Cfactor}
\end{equation}

\begin{figure}
\begin{center}
\includegraphics[scale=0.4]{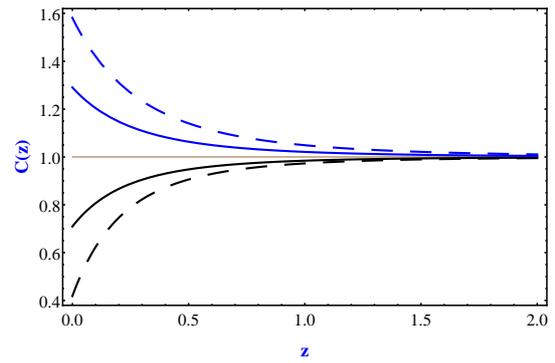}
\end{center}
\caption{\label{fig:Cofz} $ C(z) $ vs. $ z $ graphs. The color code is same as in Fig.~\ref{fig:wofz}.}
\end{figure}

\noindent
Eq. \eqref{eq:Euler2} can also be modified as

\begin{equation}
\dfrac{\partial \theta}{\partial \tau} + \mathcal{H} \theta +\dfrac{3}{2} \mathcal{H}^{2} \Omega_{m} \delta= - \vec{\nabla}.[(\vec{v}.\vec{\nabla})\vec{v}].
\label{eq:Euler3}
\end{equation}

\noindent
Here we should mention that $ C = 1 $ corresponds to the smooth dark energy with $ c_{s}^{2} = 1 $, for which $ \delta = \delta_{m} $ and for clustering dark energy $ C $ differs from $1$. In other words, the function $ C $ captures all the modifications to the equations of motion in the clustering dark energy case. So it is useful to plot such a quantity. In Fig.~\ref{fig:Cofz}, we have shown how $ C(z) $ behaves compared to $ \Lambda $CDM for the same models as in Fig.~\ref{fig:wofz} and the color code is the same. $ C(z) $ gives a rough idea about the amount of clustering. Since the acceleration of the expansion of the Universe is larger in the phantom models, the clustering is slower in the phantom models compared to the non-phantom models. More the phantom (non-phantom) behavior slower (faster) the clustering. This fact is reflected in Fig.~\ref{fig:Cofz}. Due to this reason, in phantom models, the value of $ C $ is less than $ 1 $ which is the value of $ C $ for the $ \Lambda $CDM model. Due to the same reason, opposite happens for non-phantom models. More the phantom (non-phantom) behaviour lesser (greater) the value of $ C $ from $ -1 $.  At high redshifts contributions from dark energy are negligible. So, all the models have $ C \sim 1 $ at high redshifts.

\noindent
In Fourier space Eqs. \eqref{eq:CntnTotal} and \eqref{eq:Euler3} can be written as

\begin{eqnarray}
\dfrac{\partial \delta_{\vec{k}}}{\partial \tau} +C(\tau) \theta_{\vec{k}} &=& - \int d^{3}\vec{q}_{1} \int d^{3}\vec{q}_{2} \;\; \delta_{D}^{(3)}(\vec{k}-\vec{q_{1}}-\vec{q_{2}}) \nonumber\\
&& \alpha(\vec{q_{1}}, \vec{q_{2}}) \theta_{\vec{q_{1}}} \delta_{\vec{q_{2}}},
\label{eq:CntnTotalFS}
\end{eqnarray}

\begin{eqnarray}
\dfrac{\partial \theta_{\vec{k}}}{\partial \tau} + \mathcal{H} \theta_{\vec{k}} +\dfrac{3}{2} \mathcal{H}^{2} \Omega_{m} \delta_{\vec{k}} &=& - \int d^{3}\vec{q_{1}} \int d^{3}\vec{q_{2}} \;\; \nonumber\\
&& \delta_{D}^{(3)}(\vec{k}-\vec{q_{1}}-\vec{q_{2}}) \nonumber\\
&& \beta(\vec{q_{1}},\vec{q_{2}}) \theta_{\vec{q_{1}}} \theta_{\vec{q_{2}}},
\label{eq:Euler3FS}
\end{eqnarray}

\noindent
where,

\begin{equation}
\alpha(\vec{q_{1}},\vec{q_{2}})=1+\dfrac{\vec{q_{1}}.\vec{q_{2}}}{q_{1}^2},
\label{eq:alpha}
\end{equation}

\begin{equation}
\beta(\vec{q_{1}},\vec{q_{2}})=\dfrac{(\vec{q_{1}}+\vec{q_{2}})^{2} \;\; (\vec{q_{1}}.\vec{q_{2}})}{2q_{1}^{2}q_{2}^{2}},
\label{eq:beta}
\end{equation}

\noindent
and $ \vec{k} $, $ \vec{q_{1}} $ $ \& $ $ \vec{q_{2}} $ correspond to different wave modes in the Fourier space.

\subsection{Smooth dark energy}
For smooth dark energy, there are no perturbations in the dark energy. This corresponds to the sound speed, $c_{s}^2 = 1$ of the dark energy. In this scenario, the total overdensity is same as the matter overdensity because of $ \delta_Q =0 $ i.e.

\begin{equation}
\delta_Q =0 \hspace{1.0 cm} \Longrightarrow \hspace{1.0 cm} \delta = \delta_{m}. \hspace{0.5 cm} \text{(smooth case)}
\label{eq:deltasmooth}
\end{equation}

\noindent
And also $ v_{Q}=0 $. Considering $\delta_{m}=\delta$ and $v_{m}=v$ we get the same Eqs. \eqref{eq:CntnTotalFS} and \eqref{eq:Euler3FS} in the fourier space with $ C=1 $. So,

\begin{equation}
C(\tau) = 1. \hspace{0.5 cm} \text{(smooth case)}
\label{eq:Csmooth}
\end{equation}

\noindent
Note that $C=1$ in the smooth case does not correspond to $1+(1+\omega)\dfrac{\Omega_{Q}}{\Omega_{m}} = 1$ (except for the $ \Lambda $CDM model), actually, the form of the continuity equation has no $C$ term.

\noindent
Note that, in the present analysis, we have not considered any interaction between matter and dark energy in both cases.

\section{Linear solutions} 
 
In linear regime we can neglect 2nd order (and higher orders) terms and then Eqs. \eqref{eq:CntnTotalFS} and \eqref{eq:Euler3FS} becomes

\begin{equation}
\dfrac{\partial \delta^{lin}_{\vec{k}}}{\partial \tau} +C(\tau) \theta^{lin}_{\vec{k}} = 0,
\label{eq:CntnTotalFSlinear}
\end{equation}

\begin{equation}
\dfrac{\partial \theta^{lin}_{\vec{k}}}{\partial \tau} + \mathcal{H} \theta^{lin}_{\vec{k}} +\dfrac{3}{2} \mathcal{H}^{2} \Omega_{m} \delta^{lin}_{\vec{k}}= 0,
\label{eq:Euler3FSlinear}
\end{equation}

\noindent
where superscript 'lin' stands for linear theory. In Fourier space, the evolution equation for the total density contrast can be written as 

\begin{equation}
\delta^{lin}_{\vec{k}}(\tau)=D(\tau) \delta^{in}_{\vec{k}},
\label{eq:deltaFSlinearGrowth}
\end{equation} 

\noindent
where $ D(\tau) $ is called linear growth function and $ \delta^{in}_{\vec{k}} $ is the initial density contrast at some sufficient initial time. Using above definition of the linear growth function into Eq. \eqref{eq:CntnTotalFSlinear} we get 

\begin{equation}
\theta^{lin}_{\vec{k}}= - \dfrac{\mathcal{H}(\tau) f(\tau)}{C(\tau)} D(\tau) \delta^{in}_{\vec{k}}.
\label{eq:thetalinear}
\end{equation}

\noindent
This is the corresponding evolution equation for the velocity field in Fourier space and f is the linear growth rate and it is defined as

\begin{equation}
f=\dfrac{d \; lnD}{d \; lna}.
\label{eq:growthf}
\end{equation}

\noindent 
Taking derivative of Eq. \eqref{eq:CntnTotalFSlinear} and putting Eq. \eqref{eq:Euler3FSlinear} into it (and using the definition in Eq. \eqref{eq:deltaFSlinearGrowth}), we get evolution equation for the linear growth function as

\begin{equation}
\dfrac{d^{2} D}{d N^{2}} + \bigg[ \frac{1}{2} (1-3 w \Omega_{Q}) - \frac{d lnC}{d lna} \bigg] \dfrac{d D}{d N} - \frac{3}{2} \Omega_{m} C D = 0,
\label{eq:growthFunction}
\end{equation}

\noindent
where $N=lna$ is the e-fold. Note that for the solution of Eq. \eqref{eq:growthFunction}, we only consider growing mode solutions throughout this paper. For the constant equation of state of dark energy, the growing solution can be obtained analytically both for smooth and clustering dark energy and the solution can be expressed in terms of hyper-geometric functions (S. Anselmi 2014, E. Sefusatti 2011). But for a varying equation of state for dark energy, we have to solve Eq. \eqref{eq:growthFunction} numerically.

\noindent
Now similar to the Eq. \eqref{eq:deltaFSlinearGrowth}, in Fourier space we can define individual linear growth functions for matter ($ \delta_{m} $) and dark energy ($ \delta_{Q} $) as

\begin{eqnarray}
\delta_{m_{\vec{k}}}^{lin}(\tau)=D_{m}(\tau) \delta^{in}_{\vec{k}}, \nonumber\\
\delta_{Q_{\vec{k}}}^{lin}(\tau)=D_{Q}(\tau) \delta^{in}_{\vec{k}}.
\label{eq:deltaFSlinearGrowthIndividual}
\end{eqnarray}

\begin{center}
\begin{figure*}[!h]
\begin{tabular}{c@{\quad}c}
\epsfig{file=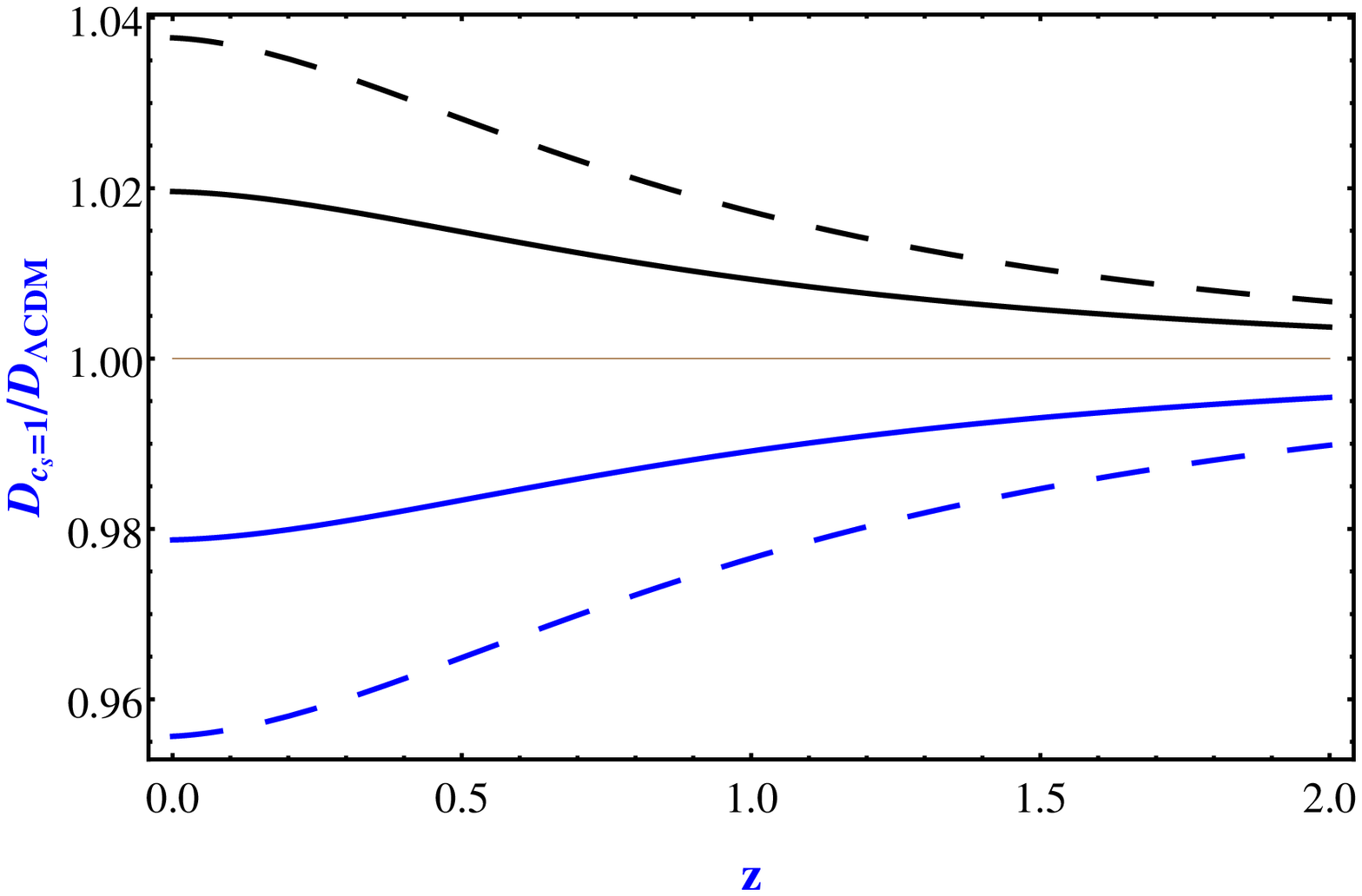,width=7.5 cm}
\epsfig{file=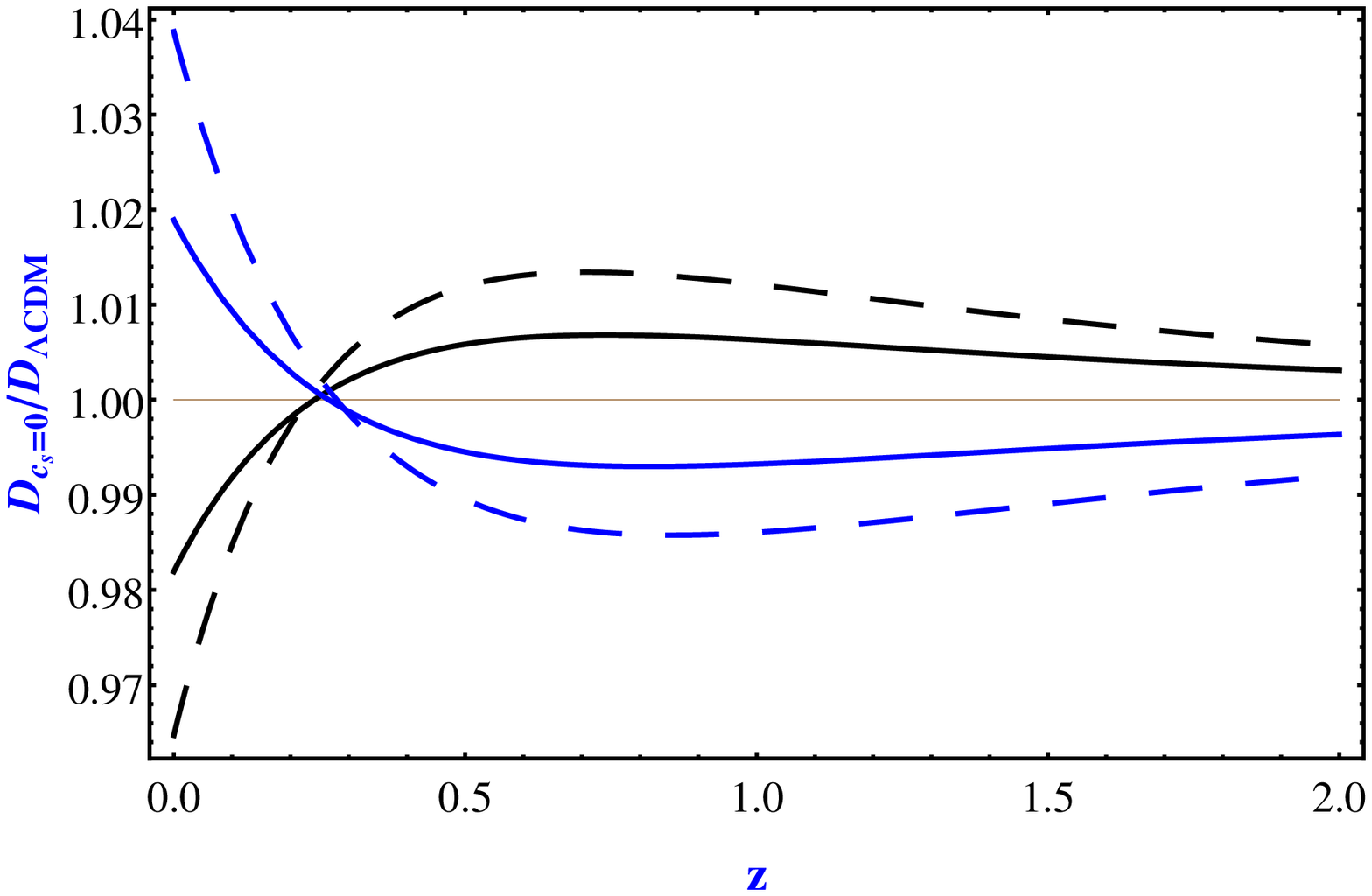,width=7.5 cm}\\
\epsfig{file=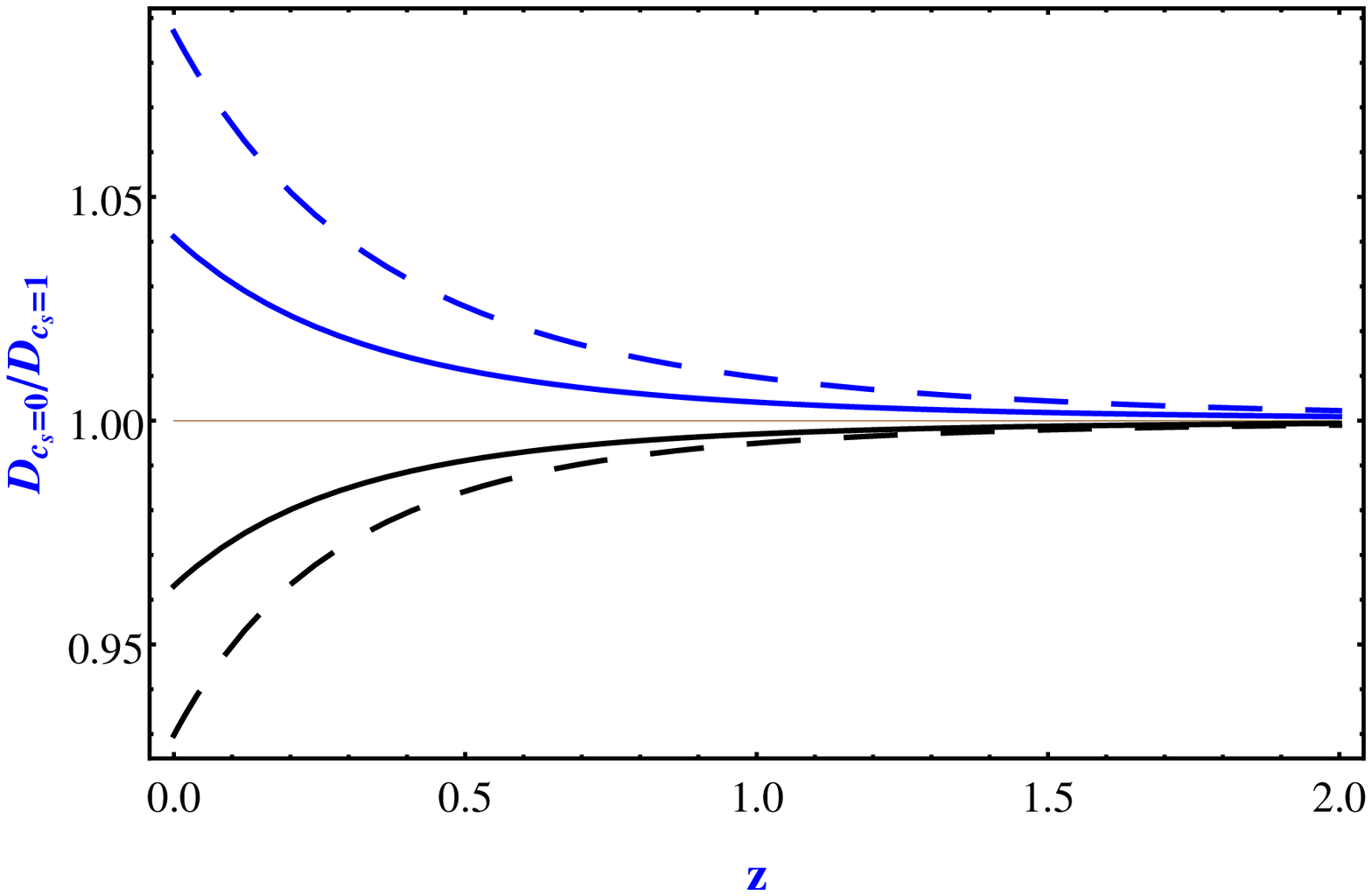,width=7.5 cm}
\epsfig{file=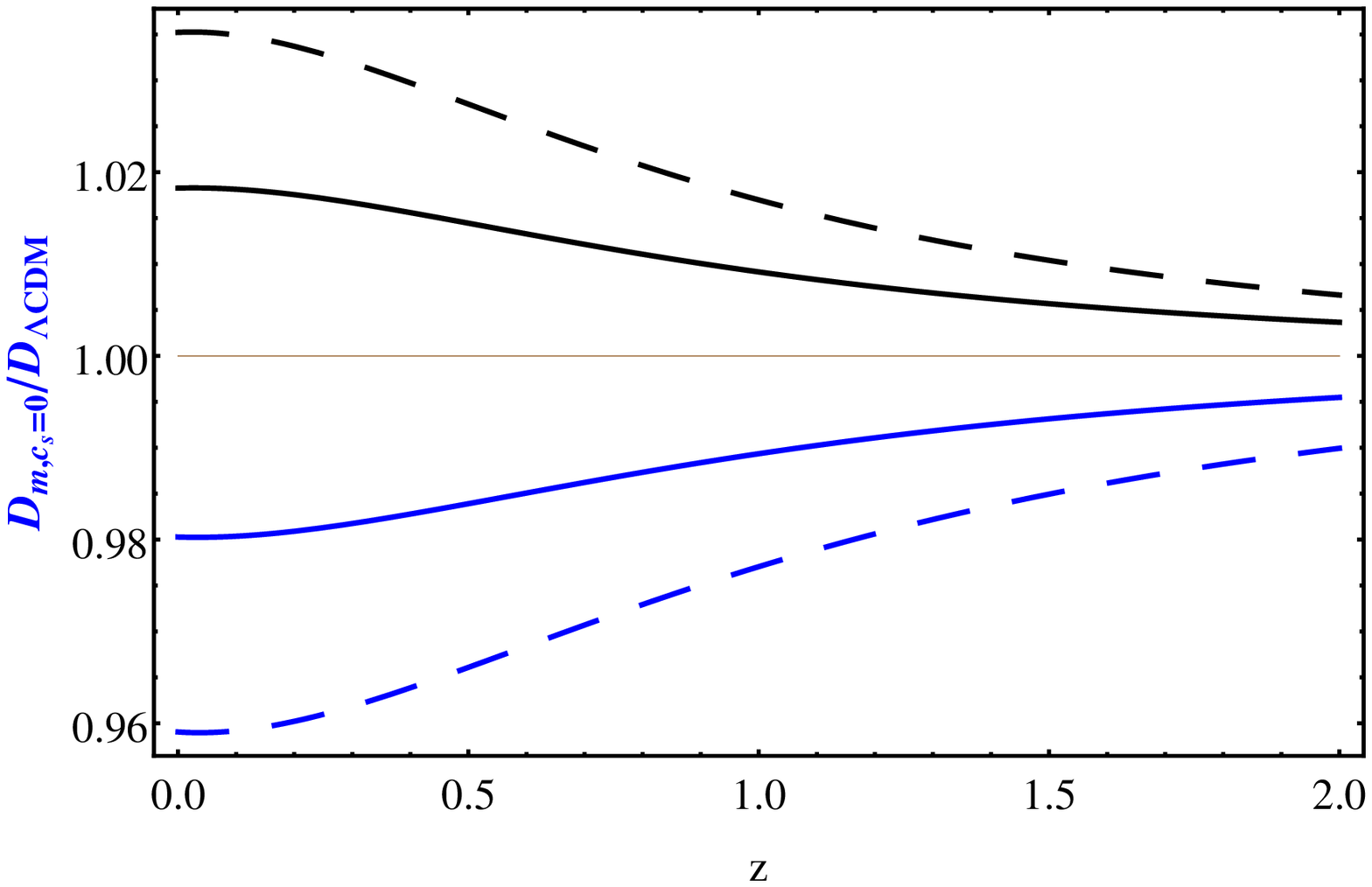,width=7.5 cm}\\
\epsfig{file=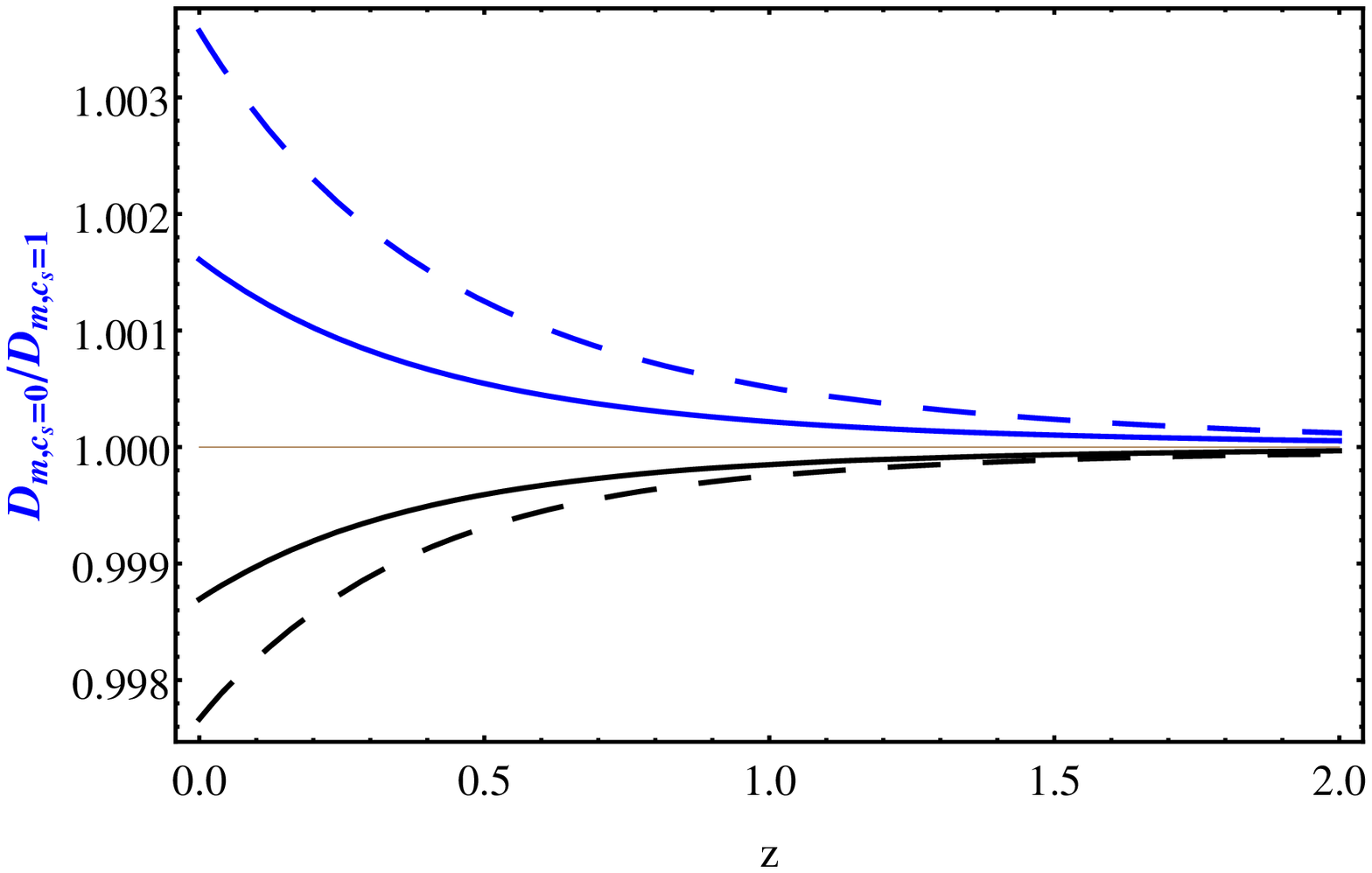,width=7.5 cm}
\end{tabular}
\caption{\label{fig:DLcmp} Comparison in $ D(z) $ between different cases as function of $ z $. Different panels have different comparison plots respectively. The color code is same as in Fig.~\ref{fig:wofz}.
}
\end{figure*}
\end{center}

\noindent
Using Eqs. \eqref{eq:CntnTotalFSlinear}, \eqref{eq:Euler3FSlinear} $\&$ \eqref{eq:deltaFSlinearGrowthIndividual} with linearised Eqs. \eqref{eq:CntnMatter} $ \& $ \eqref{eq:CntnDE} (in Fourier space), we get the relation between individual and total linear growth function as

\begin{eqnarray}
\dfrac{d D_{m}}{d N}=\dfrac{1}{C} \dfrac{d D}{d N}, \nonumber\\
\dfrac{d D_{Q}}{d N} - 3 \omega D_{Q} = \dfrac{1+\omega}{C} \dfrac{d D}{d N}.
\label{eq:GrowthFindividual}
\end{eqnarray}

\begin{figure}
\begin{center}
\includegraphics[scale=0.3]{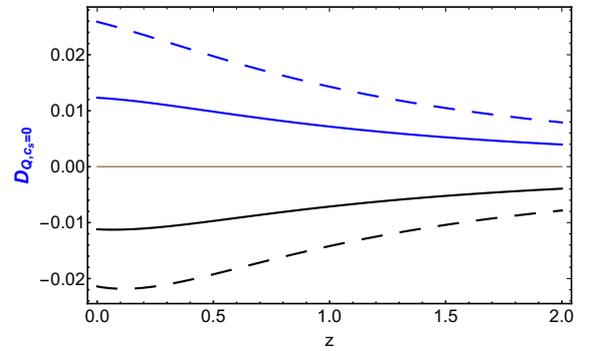}
\end{center}
\caption{\label{fig:dQcs0} $ D_{Q} $ vs. $ z $ graphs for $c_{s}^{2}=0$. The color code is same as in Fig.~\ref{fig:wofz}.}
\end{figure}

\noindent
From above equations, we can compute individual growth functions both for matter and dark energy from the total one. From above equations, it is clear that for smooth dark energy ($ C =1 $), growth functions for matter are equal to the total one up to some arbitrary constant. To fix the initial condition, we know that in the early Universe, the dark energy was negligible and the Universe was dominated by matter only. Hence we fix the initial condition at $z=1000$. The initial condition is given by $ D/a \rightarrow 1 $, $ D_{m}/a \rightarrow 1 $, and $ D_{Q}/a \rightarrow 0 $ respectively for $a \rightarrow 0$.

In Fig.~\ref{fig:DLcmp}, we show how individual and total growth functions $ D $ evolve with respect to redshift for different cases. Here one can see both the parameter $w_{0}$ and $w_{a}$ play the important role for clustering quintessence to differ from $\Lambda$CDM as well as from smooth quintessence case. The color code is the same as in Fig.~\ref{fig:wofz}.

In the top-left panel of Fig.~\ref{fig:DLcmp}, we compare the total growth functions from the $\Lambda$CDM one for the smooth dark energy models. Since there is no dark energy clustering, dark energy only effects through the background evolution. This effect corresponds to the deviations up to $2-4\%$ at low redshifts for the considered models. Since at the background level, in the non-phantom models, dark energy dominates (over matter) earlier i.e. the accelerated era starts earlier, the clustering in matter slows down in the non-phantom models compared to the phantom models. More the non-phantom (phantom) behavior slower (faster) the clustering in the matter.

In the top-right panel of Fig.~\ref{fig:DLcmp}, we compare the total growth functions from the $\Lambda$CDM one for the clustering dark energy models. Unlike the previous case, here the clustering in dark energy is present along with the clustering in the matter. We can see this dark energy clustering effect through the function $ C(z) $ from Fig.~\ref{fig:Cofz}. Since, $ C \sim 1 $ at large redshifts, we can see the same behavior as in the previous case (i.e. the top-left panel) at large redshifts. The clustering of dark energy changes the behavior in total growth function at smaller redshifts. As mentioned before, the dark energy clustering is slower in the phantom models due to the faster acceleration compared to the non-phantom models (can be seen from  Fig.~\ref{fig:Cofz}). Whereas at the background level phantom models have the faster clustering in matter compared to the $\Lambda$CDM or non-phantom models. So, these are the two opposite effects. At smaller redshifts, hindering of the dark energy clustering wins over the enhancement of the matter clustering in the phantom models compared to the $\Lambda$CDM model. Due to the same reason, the opposite happens for the non-phantom models. So, we see the crossing in the graphs at lower redshifts both for phantom and non-phantom models.

In the middle-left panel of Fig.~\ref{fig:DLcmp}, we have compared total growth for the clustering dark energy from the smooth case for the same model. This plot gives an idea about how only the extra clustering in dark energy behaves differently for different models. We can see this fact because the clustering in the matter is canceled out by taking the ratio between clustering and smooth cases. From the previous discussion, we know the fact that the dark energy clustering is faster in the non-phantom models since the acceleration is slower compared to the $\Lambda$CDM or the phantom models. The opposite happens for the phantom models. This fact is reflected in this panel.

In the middle-right panel of Fig.~\ref{fig:DLcmp}, we have compared the matter growth function for the clustering dark energy models from the $\Lambda$CDM one. Although the clustering in dark energy is present in the clustering dark energy models, by plotting the matter growth function we are considering the clustering in the matter only. On the other hand, for the smooth dark energy models, there is only the clustering in the matter. So, basically, the top-left and the middle-right panels have the similar behavior.

In the bottom panel of Fig.~\ref{fig:DLcmp}, we have compared the matter growth function between clustering and smooth cases for the same dark energy model respectively. By this plot, we are basically separating out the effect of dark energy only at background level. At background level, we know from the previous discussion that non-phantom models have faster clustering in matter compared to the $\Lambda$CDM or the phantom models. The opposite happens for the phantom models. This fact is reflected in this panel.

Note that in this paper, all the perturbation equations are valid for sub-Hubble scales. Although on the sub-Hubble scales, the dark energy perturbations are negligible for smooth case, on super-Hubble or near Hubble scales the dark energy perturbations are not negligible even for the smooth case. So, the perturbations on super-Hubble or near Hubble scales are beyond our discussions of this paper. Now in the sub-Hubble scale and in the linear regime, we have an accurate description of the difference between the smooth case and the clustering case. This has been discussed in Figs.~\ref{fig:Cofz} and~\ref{fig:DLcmp}. For the clustering case, we can see how the individual components (matter or dark energy) cluster accordingly through the perturbation Eq. \eqref{eq:GrowthFindividual}. To see this, in Fig.~\ref{fig:dQcs0}, we have plotted contribution of perturbation from dark energy component only ($D_{Q}$) for clustering case. We know, for the smooth case, $D_{Q}=0$. But, from Fig.~\ref{fig:dQcs0}, we can see that for the clustering case dark energy perturbation is non-zero at smaller redshifts. This result is almost similar on non-linear scales. Only a bit extra effect comes on small scales at smaller redshifts only according to the bottom-most panel of the Fig.~\ref{fig:DLcmp}. This will be discussed later in subsection (5.3) and section 6.

Using definition of the linear growth rate (Eq. \eqref{eq:growthf}) into Eq. \eqref{eq:growthFunction} we get its evolution equation as

\begin{equation}
\dfrac{d f}{d ln a} +f^{2} + [2+\dfrac{d lnH}{d lna}-\dfrac{d lnC}{d lna}] f = - \dfrac{d lnH}{d lna},
\label{eq:fEvolution}
\end{equation}

\noindent
where growth rate, $f$ is defined as $f = \dfrac{d lnD}{d lna}$.

\section{Non-linear solutions} 

\subsection{Evolution of the non-linear propagator and power spectrum}

We define a quantity $ \eta $ which can be related to time is given by (S. Anselmi 2012, S. Anselmi 2014, E. Sefusatti 2011)

\begin{equation}
\eta = log\Big{[}\frac{D}{D_{in}}\Big{]}.
\label{eq:defnEta}
\end{equation}

\noindent
Using this definition Eqs. \eqref{eq:CntnTotalFS} and \eqref{eq:Euler3FS} can be rewritten as (E. Sefusatti 2011)

\begin{eqnarray}
\dfrac{\partial \delta_{\vec{k}}}{\partial \eta} - \Theta_{\vec{k}} &=& \int d^{3}\vec{q}_{1} \int d^{3}\vec{q}_{2} \;\; \delta_{D}^{(3)}(\vec{k}-\vec{q_{1}}-\vec{q_{2}}) \nonumber\\
&& \frac{\alpha(\vec{q_{1}}, \vec{q_{2}})}{C(\eta)} \Theta_{\vec{q_{1}}} \delta_{\vec{q_{2}}},
\label{eq:deltaEvlvNL}
\end{eqnarray}

\begin{eqnarray}
\dfrac{\partial \Theta_{\vec{k}}}{\partial \eta} - \Theta_{\vec{k}} -\dfrac{f_{-}}{f_{+}^{2}} (\Theta_{\vec{k}} - \delta_{\vec{k}}) &=& \int d^{3}\vec{q_{1}} \int d^{3}\vec{q_{2}} \;\; \nonumber\\
&& \delta_{D}^{(3)}(\vec{k}-\vec{q_{1}}-\vec{q_{2}}) \nonumber\\
&& \frac{\beta(\vec{q_{1}},\vec{q_{2}})}{C(\eta)} \Theta_{\vec{q_{1}}} \Theta_{\vec{q_{2}}},
\label{eq:thetaEvlvNL}
\end{eqnarray}

\noindent
where we have defined

\begin{equation}
\Theta_{\vec{k}} = -\frac{C}{\mathcal{H} f} \theta_{\vec{k}}.
\label{eq:defnCaptheta}
\end{equation}

\noindent
Now we define a doublet $ \psi_{a} $ ($ a = 1,2 $) which is given by

\begin{eqnarray}
\begin{bmatrix}
\psi_{1}(\vec{k},\eta) \\      \psi_{2}(\vec{k},\eta) \end{bmatrix} \equiv e^{-\eta} \begin{bmatrix}
\delta_{\vec{k}}(\eta) \\      \Theta_{\vec{k}}(\eta) \end{bmatrix}.
\label{eq:doublet}
\end{eqnarray}

\noindent
Using this definition we can write Eqs. \eqref{eq:deltaEvlvNL} and \eqref{eq:thetaEvlvNL} in a matrix form given by (S. Anselmi 2014)

\begin{eqnarray}
&& \partial_{\eta} \psi_{a}(\vec{k},\eta) = - \Omega_{ab} (\eta) \psi_{b}(\vec{k},\eta) \nonumber\\
&& + \frac{e^{\eta}}{C(\eta)} \gamma_{abc}(\vec{k},-\vec{p},-\vec{q}) \psi_{b}(\vec{p},\eta) \psi_{c}(\vec{q},\eta),
\label{eq:doubleEvolution}
\end{eqnarray}

\noindent
where repeated indices are summed over and  we also introduce a vertex matrix, $ \gamma_{abc}(\vec{k},\vec{p},\vec{q}) $ ($ a,b,c = 1,2 $) which has only non-vanishing, independent elements given by

\begin{eqnarray}
&& \gamma_{121}(\vec{k},\vec{p},\vec{q}) = \frac{1}{2} \int d^{3}\vec{p} \int d^{3}\vec{q} \;\; \delta_{D}^{(3)}(\vec{k}+\vec{p}+\vec{q}) \alpha(\vec{p}, \vec{q}), \nonumber\\
&& \gamma_{222}(\vec{k},\vec{p},\vec{q}) = \int d^{3}\vec{p} \int d^{3}\vec{q} \;\; \delta_{D}^{(3)}(\vec{k}+\vec{p}+\vec{q}) \beta(\vec{p}, \vec{q}), \nonumber\\
&& \gamma_{112}(\vec{k},\vec{p},\vec{q}) = \gamma_{121}(\vec{k},\vec{q},\vec{p}),
\label{eq:gammas}
\end{eqnarray}

\noindent
and $ \Omega_{ab} $ matrix is given by

\begin{eqnarray}
\Omega =  \begin{bmatrix}
    1 &\hspace{0.5 cm} -1\\      \frac{f_{-}}{f^{2}} &\hspace{0.5 cm} -\frac{f_{-}}{f^{2}}\      \end{bmatrix},
\label{eq:OmegaMatrix}    
\end{eqnarray}

\noindent
where $f_{-}$ is given by $f_{-} = \dfrac{d lnH}{d lna}=-\dfrac{3}{2} (1+w \Omega_{Q})$.

\begin{center}
\begin{figure*}[!h]
\begin{tabular}{c@{\quad}c}
\epsfig{file=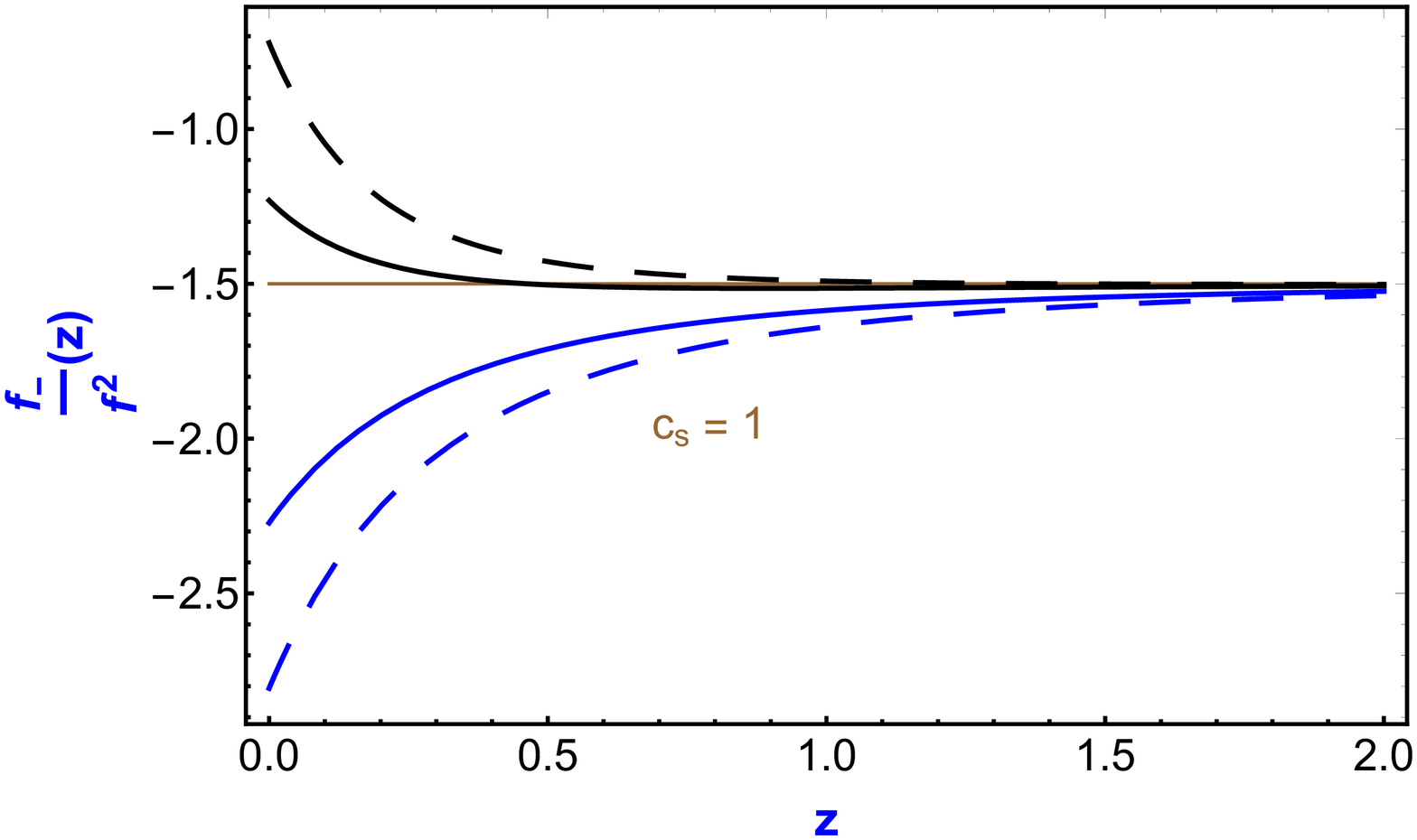,width=7.5 cm}
\epsfig{file=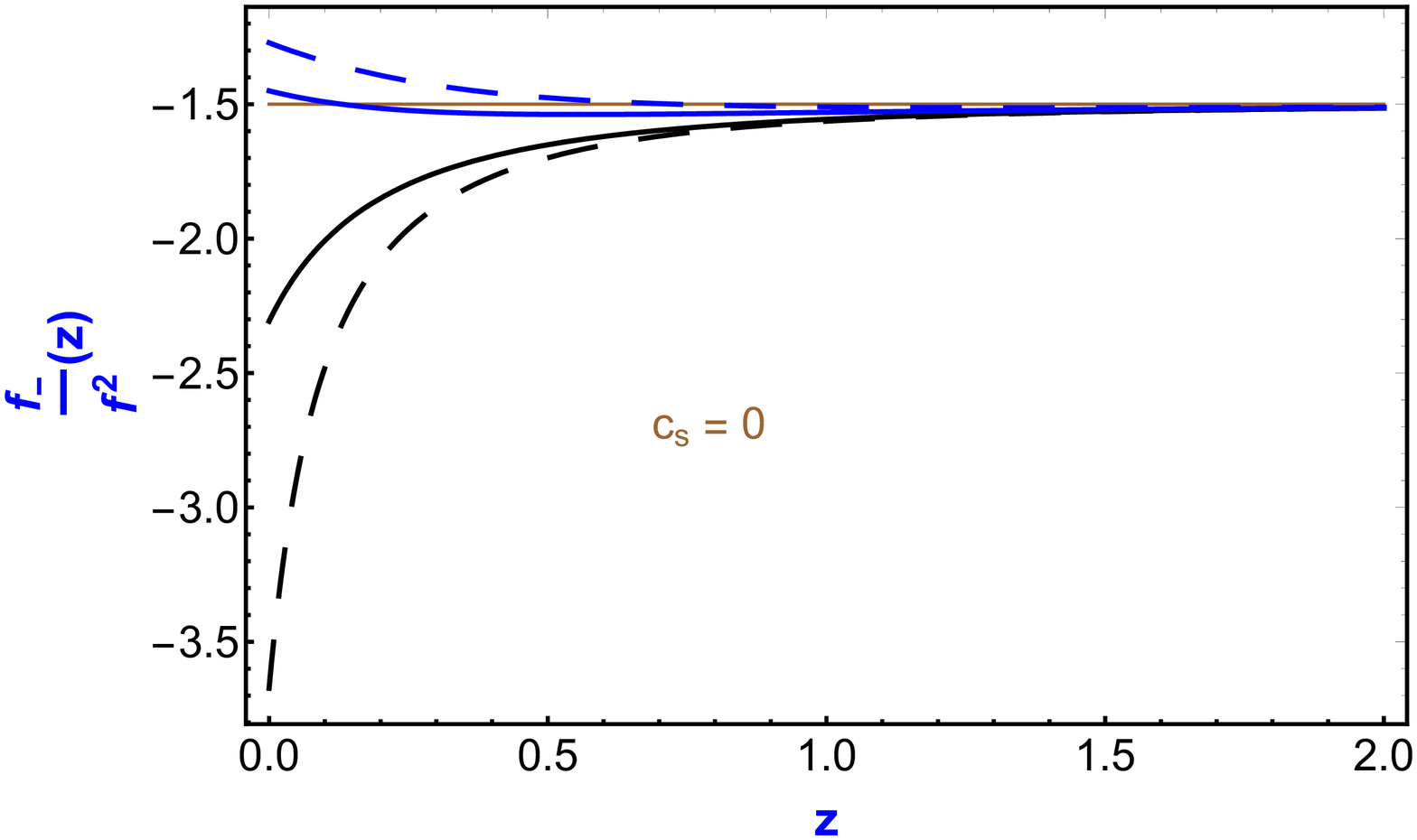,width=7.5 cm}
\end{tabular}
\caption{\label{fig:fminusbyfplus} $ f_{-}/f^{2}(z) $ vs. $ z $ plots. The color code is same as in Fig.~\ref{fig:wofz}.}
\end{figure*}
\end{center}

\noindent
Now, as $ \Omega $ matrix is time dependent it is difficult to directly implement the resummation technique. But in our case we can use the approximation $ \frac{f_{-}}{f^{2}} \approx -3/2 $ which is a reasonable assumption as one can see from Fig.~\ref{fig:fminusbyfplus}. In this figure we plot $ f_{-}/f^{2}(z) $ with respect to redshift $z$ for the same models as in Fig.~\ref{fig:wofz} both for smooth and clustering cases. One can see that for most of the cases, $ f_{-}/f^{2} $ tends to -3/2 for all redshifts except for very low redshifts. Here we should mention that for non-phantom models this approximation holds better while it worsens for phantom models in the clustering dark energy. It is also interesting to see that for phantom models this approximation holds comparatively better for smooth dark energy compared to clustering dark energy (keeping same $ w_{0} $ and $ w_{a} $ values). In our subsequent calculations, we use the approximation

\begin{eqnarray}
\Omega \approx  \begin{bmatrix}
    1 &\hspace{0.5 cm} -1\\      -\frac{3}{2} &\hspace{0.7 cm}   \frac{3}{2}\      \end{bmatrix}.
\label{eq:OmegaMatrixAppx}
\end{eqnarray}

\noindent
The linear solution $ \psi_{a}^{L} $ can be expressed as

\begin{equation}
\psi_{a}^{L} (\vec{k},\eta) = g_{ab}(\eta,\eta') \psi_{b}^{L} (\vec{k},\eta'),
\label{eq:doubletLinearSol}
\end{equation}

\noindent
where $ g_{ab} $ is the retarded linear propagator which obeys

\begin{equation}
(\delta_{ab} \partial_{\eta} + \Omega_{ab}) g_{bc}(\eta,\eta') = \delta_{ac} \delta_{D} (\eta - \eta')
\label{eq:gabEvolution}
\end{equation}

\noindent
From Eq. \eqref{eq:gabEvolution} and using the approximation \eqref{eq:OmegaMatrixAppx}, one can easily calculate the linear propagator which is given by (S. Anselmi 2012, S. Anselmi 2014)

\begin{equation}
g_{ab}(\eta,\eta') = [B+A e^{-5/2(\eta-\eta')}] \theta(\eta-\eta'),
\label{eq:gabSol}
\end{equation}

\noindent
where $ \theta $ is the step function and $ A $ and $ B $ are given by

\begin{eqnarray}
A = \frac{1}{5} \begin{bmatrix}
    2 &\hspace{0.2 cm} -2\\      -3 &\hspace{0.2 cm}   3\      \end{bmatrix} \hspace{0.5 cm} and \hspace{0.5 cm}
B = \frac{1}{5} \begin{bmatrix}
    3 &\hspace{0.2 cm} 2\\      3 &\hspace{0.2 cm}   2\      \end{bmatrix}.
\label{eq:AandBmatrix}
\end{eqnarray}

\noindent
Using above equations one can see the initial fields $ \psi_{a} $ are proportional to

\begin{eqnarray}
u_{a} = \begin{bmatrix}
    1\\1\      \end{bmatrix} \hspace{1 cm} and \hspace{1 cm}
v_{a} = \begin{bmatrix}
    1\\      -3/2\      \end{bmatrix} ,  
\label{eq:uvdoublet}
\end{eqnarray}

\noindent
which correspond to growing and decaying mode respectively. Since we are only considering the growing mode solution, we shall only use the $u$ vector in the next sections. Subsequently, we can compute the linear power spectrum as 

\begin{equation}
P_{ab}^{L} (\vec{k},\eta,\eta') = g_{ac}(\eta,\eta_{in}) g_{bd}(\eta',\eta_{in}) P_{cd}^{0} (\vec{k},\eta_{in},\eta_{in}),
\label{eq:PSlinear}
\end{equation}

\noindent
where $ P_{ab}^{0} (\vec{k},\eta_{in},\eta_{in}) \simeq P_{in}(k) u_{a} u_{b} $, where $ P_{in}(k) $ be the initial density power spectrum. Till now we have computed all the important linear quantities like growth function, propagator and power spectrum; our next task is to compute full non-linear quantities, mainly non-linear density power spectrum.
\\
\noindent
To start with we can write down the full non-linear propagator and power spectrum as (see (M. Crocce 2006a, S.~Matarrese 2007, S.~Anselmi 2011b,S. Anselmi 2012) and in diagrammatic representation see (M. Crocce 2006c)) 

\begin{eqnarray}
&& G_{ab}(k;\eta,\eta')= g_{ab}(\eta-\eta') + \int_{\eta'}^{\eta} ds_{1} \int_{\eta'}^{s_{1}} ds_{2} \nonumber\\
&& g_{ac}(\eta-s_{1}) \Sigma_{cd}(k;s_{1},s_{2}) G_{db}(k;s_{2},\eta')
\label{eq:GabDifferentWay}
\end{eqnarray}

\noindent
and

\begin{eqnarray}
&& P_{ab}(k;\eta,\eta')=G_{ac}(k;\eta,\eta_{in})G_{bd}(k;\eta',\eta_{in})P^{0}_{cd}(k;\eta_{in},\eta_{in})\nonumber\\
&&+\int_{0}^{\eta} ds_{1} \int_{0}^{\eta'} ds_{2} \nonumber\\
&& G_{ac}(k;\eta,s_{1}) G_{bd}(k;\eta',s_{2}) \Phi_{cd}(k;s_{1},s_{2})
\label{eq:PSdifferentWay}
\end{eqnarray}

\noindent
respectively, where $ g_{ab} $ and $ P^{0} $ are the non-interacting free parts (i.e. linear part which we have discussed before) and $ \Sigma_{ab} $ and $ \Phi_{ab} $ are the interacting ones (in language of quantum field theory, these are called one-particle irreducible functions). To get the detailed informations about these one-particle irreducible functions see (M. Crocce 2006a, S.~Matarrese 2007, S.~Anselmi 2011b,S. Anselmi 2012, S. Anselmi 2014, M. Crocce 2006c). 
\\
\noindent
Taking the $ \eta $-derivative of Eqs. \eqref{eq:GabDifferentWay} and \eqref{eq:PSdifferentWay} one can write the evolution equations for the full non-linear propagator and the power spectrum as (M. Crocce 2006a, S. Anselmi 2012, M. Crocce 2006c)

\begin{eqnarray}
&& \partial_{\eta} G_{ab}(k;\eta,\eta') = \delta_{ab} \delta_{D}(\eta-\eta')-\Omega_{ac} G_{cb}(k;\eta,\eta') \nonumber\\
&& + \int_{\eta'}^{\eta} ds \Sigma_{ac}(k;\eta,s) G_{cb}(k;s,\eta'),
\label{eq:GabEvolve}
\end{eqnarray}

\noindent
and

\begin{eqnarray}
&& \partial_{\eta} P_{ab}(k;\eta,\eta') = -\Omega_{ac} P_{cb}(k;\eta,\eta')-\Omega_{bc} P_{ac}(k;\eta,\eta')\nonumber\\
&&+\Big{[} \Delta G_{ac}(k;\eta,\eta') G_{bd}(k;\eta,\eta') \nonumber\\
&& +G_{ac}(k;\eta,\eta') \Delta G_{bd}(k;\eta,\eta') \Big{]} P^{0}(k) u_{c} u_{d}\nonumber\\
&&+\int_{\eta_{in}}^{\eta} ds_{1} \Big{[}\Phi_{ac}(k;\eta,s_{1}) G_{bc}(k;\eta,s_{1}) \nonumber\\
&& +G_{ac}(k;\eta,s_{1}) \Phi_{cb}(k;s_{1},\eta)\Big{]}\nonumber\\
&&+\int_{\eta_{in}}^{\eta} ds_{1} \int_{\eta_{in}}^{\eta'} ds_{2} \Phi_{cd}(k;s_{1},s_{2}) \nonumber\\
&& \Big{[} \Delta G_{ac}(k;\eta,s_{1}) G_{bd}(k;\eta,s_{2})\nonumber\\
&&+G_{ac}(k;\eta,s_{1}) \Delta G_{bd}(k;\eta,s_{2}) \Big{]},
\label{eq:PSEvolve}
\end{eqnarray}

\noindent
respectively, where

\begin{equation}
\bigtriangleup G_{ab}(k;\eta,\eta') = \int_{\eta'}^{\eta} ds \Sigma_{ad}(k;\eta,s) G_{db}(k;s,\eta')
\label{eq:DeltaGab}
\end{equation}

\noindent
To get non-linear power spectrum, first we have to solve Eq. \eqref{eq:GabEvolve} and then using this solution of the non-linear propagator, we have to solve Eq. \eqref{eq:PSEvolve}. In the next sub-section, we will discuss how to solve these evolution equations.

\subsection{Propagator and the power spectrum in the mildly non-linear regime}

In this subsection, we exactly follow the same calculation as in (S. Anselmi 2014). So, we present this subsection in a way that it summarizes the calculations and main results of (S. Anselmi 2014). For details refer (S. Anselmi 2014) and for further details of the philosophy refer (M. Crocce 2006c). Eqs. \eqref{eq:GabEvolve} and \eqref{eq:PSEvolve} are integrodifferential equations which are very difficult to solve unless some approximations are considered. Fortunately, these equations simplify in both large and small k limits and one can also have an approximate common structure in both limits. Indeed, it has been shown that in the large k (or eikonal) limit, the kernel $ \bigtriangleup G_{ab} $ factorize as (S. Anselmi 2012, S. Anselmi 2014)

\begin{equation}
\bigtriangleup G_{ab}(k;\eta,\eta') \backsimeq H_{\bold{a}}(k;\eta,\eta') G_{\bold{a}b}(k;\eta,\eta'),
\label{eq:DeltaGabAppx}
\end{equation}

\noindent
where bold indices are not summed over (hereafter we will use this notation) and $ H_{a} $ has the form

\begin{equation}
H_{a}(k;\eta,\eta') = \int_{\eta'}^{\eta} ds \Sigma_{ab}^{(1)} (k;\eta,s) u_{b},
\label{eq:HaAppx}
\end{equation}

\noindent
where $ \Sigma_{ab}^{(1)} $ is the 1-loop approximation to the full $ \Sigma_{ab} $ and it is given by (S. Anselmi 2014)

\begin{eqnarray}
&& \Sigma_{ab}^{(1)} (k;\eta,\eta') = 4 \frac{e^{\eta +\eta'}}{C(\eta)C(\eta')} \int d^{3} \vec{q} \gamma_{acd}(\vec{k},-\vec{q},\vec{q}-\vec{k}) u_{c} \nonumber\\
&& P_{in}(q) u_{e} \gamma_{feb}(\vec{k}-\vec{q},\vec{q},-\vec{k}) g_{df}(\eta,\eta').
\label{eq:Sigma1loop}
\end{eqnarray}

\noindent
One can also see that for the small k limit i.e. in the linear regime, the same factorization holds up to a very good approximation (S. Anselmi 2012, S. Anselmi 2014). Therefore we can assume same structure by interpolating between the small k and large k regimes i.e. Eq. \eqref{eq:DeltaGabAppx} also holds for intermediate values of k, which leads to a simplified equation for the non-linear propagator given by

\begin{eqnarray}
\partial_{\eta} \bar{G}_{ab}(k;\eta,\eta') &=& \delta_{ab} \delta_{D}(\eta-\eta')-\Omega_{ac} \bar{G}_{cb}(k;\eta,\eta') \nonumber\\
&& + H_{\bold{a}}(k;\eta,\eta') \bar{G}_{\bold{a}b}(k;\eta,\eta'),
\label{eq:GevolutionAppx}
\end{eqnarray}

\noindent
where we denote $ \bar{G} $ to indicate non-linear propagator for intermediate k regime. In fact, the solution of the above equation is exact in the large k limit and reduces to the 1-loop propagator for small k limit. The solution of this equation in the large k limit becomes (S. Anselmi 2014)

\begin{eqnarray}
&& \bar{G}_{\bold{a}b}(k;\eta,\eta') \rightarrow G_{ab}^{eik}(k;\eta,\eta') \nonumber\\
&& = g_{ab}(\eta,\eta') \exp[-\frac{1}{2} k^2 \sigma_{v}^2 \mathcal{I}^{2}(\eta,\eta')],
\label{eq:GeikAppx}
\end{eqnarray}

\noindent
where superscript 'eik' stands for the eikonal limit i.e. large k limit. Here $ \sigma_{v}^2 $ and $ \mathcal{I} $ is given by (S. Anselmi 2012, S. Anselmi 2014)

\begin{equation}
\sigma_{v}^2 = \frac{1}{3} \int d^{3} \vec{q}  \frac{P_{in}(q)}{q^2},
\label{eq:sigmav2}
\end{equation}

\begin{equation}
\mathcal{I}(\eta,\eta') = \int_{\eta'}^{\eta} ds \frac{e^s}{C(s)} = \frac{D_{m}(\eta)-D_{m}(\eta')}{D_{in}},
\label{eq:curlyI}
\end{equation}

\noindent
respectively, where $ D_{in} = D(\eta_{in}) $ is the growth factor at initial time $ \eta_{in} $. Using above approximate solution for the non-linear propagator along with Eq. \eqref{eq:PSEvolve} (taking same time arguments i.e. $ \eta = \eta' $), in a similar fashion we can also write approximate equations for the power spectrum in the intermediate k regime as (S. Anselmi 2012, S. Anselmi 2014)

\begin{eqnarray}
&& \partial_{\eta} \bar{P}_{ab}(k;\eta) = -\Omega_{ac}(\eta) \bar{P}_{cb}(k;\eta)-\Omega_{bc}(\eta) \bar{P}_{ac}(k;\eta) \nonumber\\
&& +H_{\bold{a}}(k;\eta,\eta_{in}) \bar{P}_{\bold{a}b}(k;\eta)+H_{\bold{b}}(k;\eta,\eta_{in}) \bar{P}_{a\bold{b}}(k;\eta) \nonumber\\
&&+\int_{\eta_{in}}^{\eta} ds \Big{[} \tilde{\Phi}_{ac}(k;\eta,s)\bar{G}_{bc}(k;\eta,s) \nonumber\\
&& + \bar{G}_{ac}(k;\eta,s)\tilde{\Phi}_{cb}(k;\eta,s) \Big{]},
\label{eq:PSevolutionAppx}
\end{eqnarray}

\noindent
where the expression for $ \tilde{\Phi}_{ab} $ is given by (S. Anselmi 2014)

\begin{eqnarray}
\tilde{\Phi}_{ab}(k;\eta,\eta') &=& \exp[-\frac{1}{2} k^2 \sigma_{v}^2 \mathcal{I}^{2}(\eta,\eta')] \Big{[} \Phi_{ab}^{(1)}(k;\eta,\eta') \nonumber\\
&&+F(k) P_{in}(k) u_{a} u_{b} (k^2 \sigma_{v}^2)^2 \nonumber\\
&& \frac{e^{\eta +\eta'}}{C(\eta)C(\eta')} \mathcal{I}(\eta,\eta_{in}) \mathcal{I}(\eta',\eta_{in}) \Big{]},
\label{eq:PhiTildewithFilter}
\end{eqnarray}

\noindent
where $ \Phi_{ab}^{(1)} $ is the 1-loop approximation to the full $ \Phi_{ab} $ and it is given by (S. Anselmi 2014)

\begin{eqnarray}
&& \Phi_{ab}^{(1)} (k;\eta,\eta') = 2 \frac{e^{\eta +\eta'}}{C(\eta)C(\eta')} \int d^{3} \vec{q} \gamma_{acd}(\vec{k},-\vec{q},-\vec{p}) \nonumber\\
&& u_{c} P_{in}(q) u_{e} u_{d} P_{in}(p) u_{f} \gamma_{bef}(-\vec{k},\vec{q},\vec{p}),
\label{eq:Phi1Loop}
\end{eqnarray}

\begin{center}
\begin{figure*}
\begin{tabular}{c@{\quad}c}
\epsfig{file=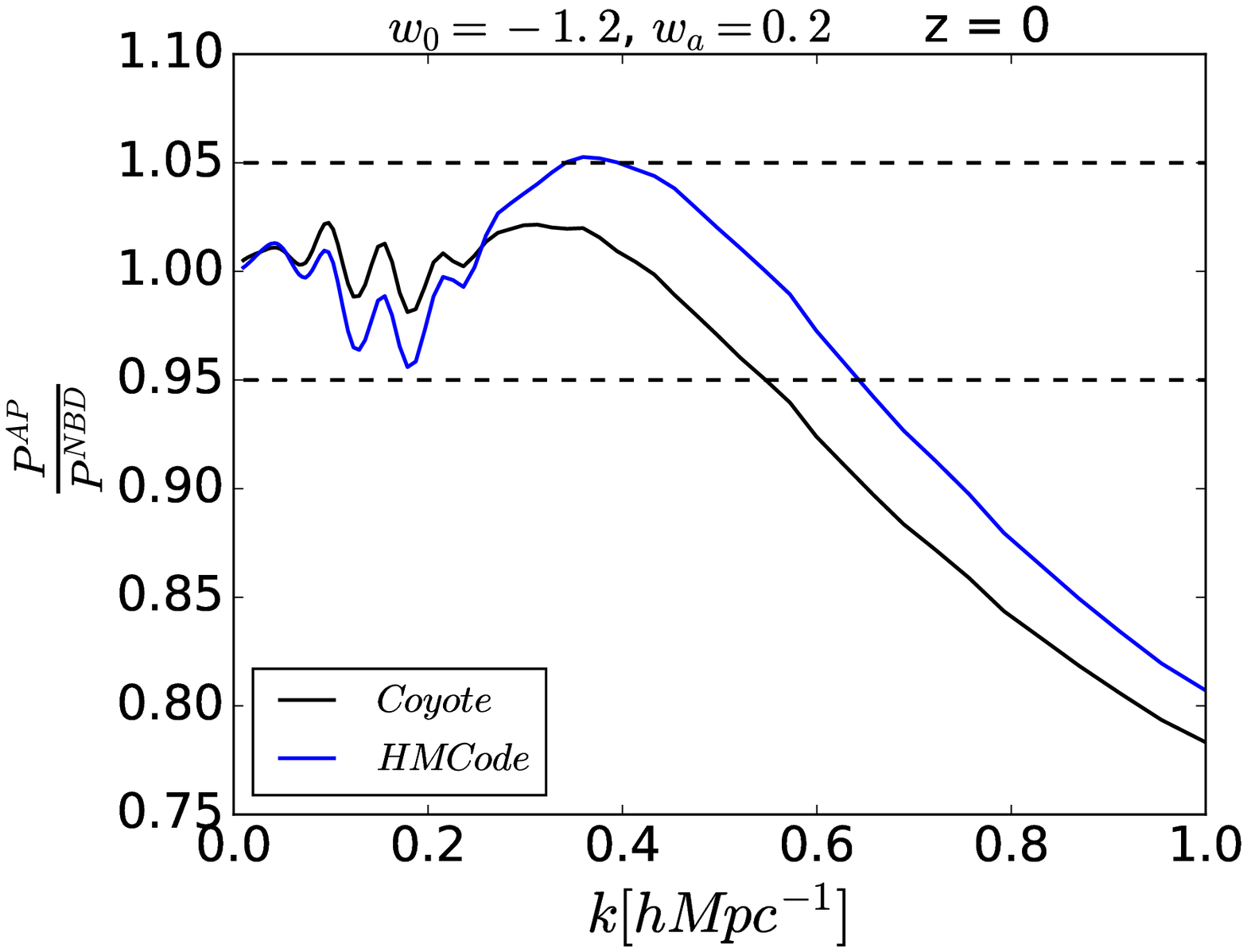,width=5.5 cm}
\epsfig{file=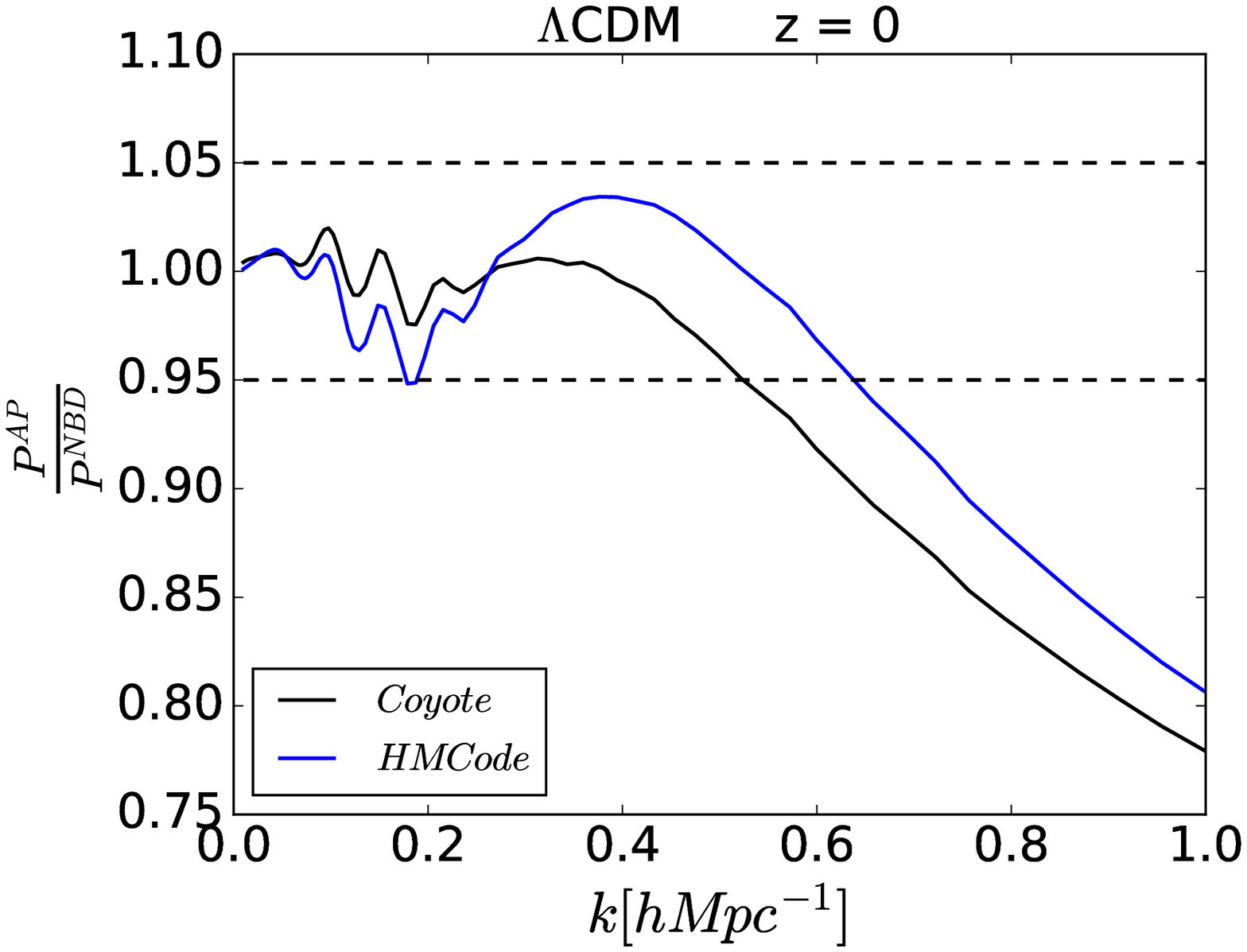,width=5.5 cm}
\epsfig{file=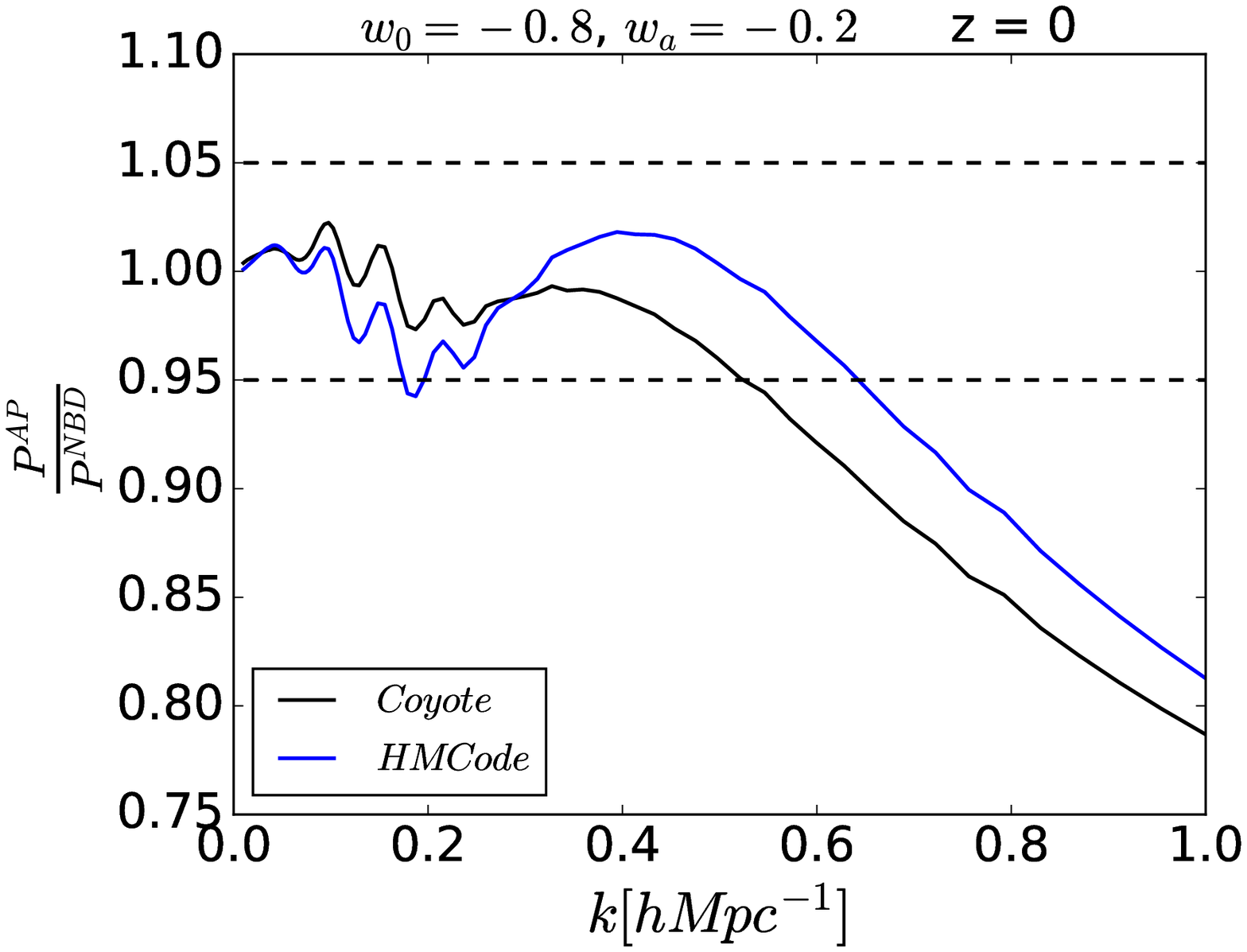,width=5.5 cm}\\
\epsfig{file=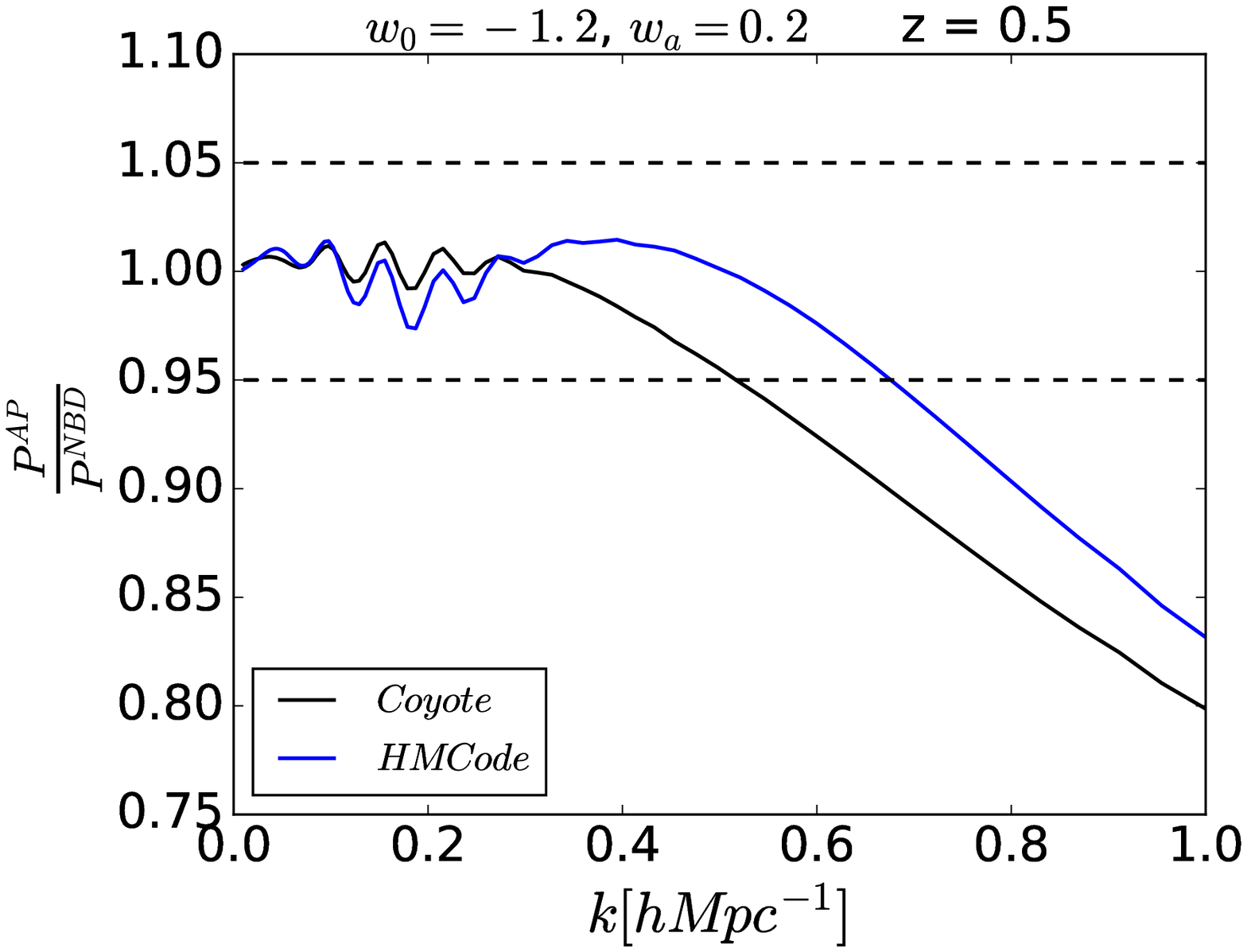,width=5.5 cm}
\epsfig{file=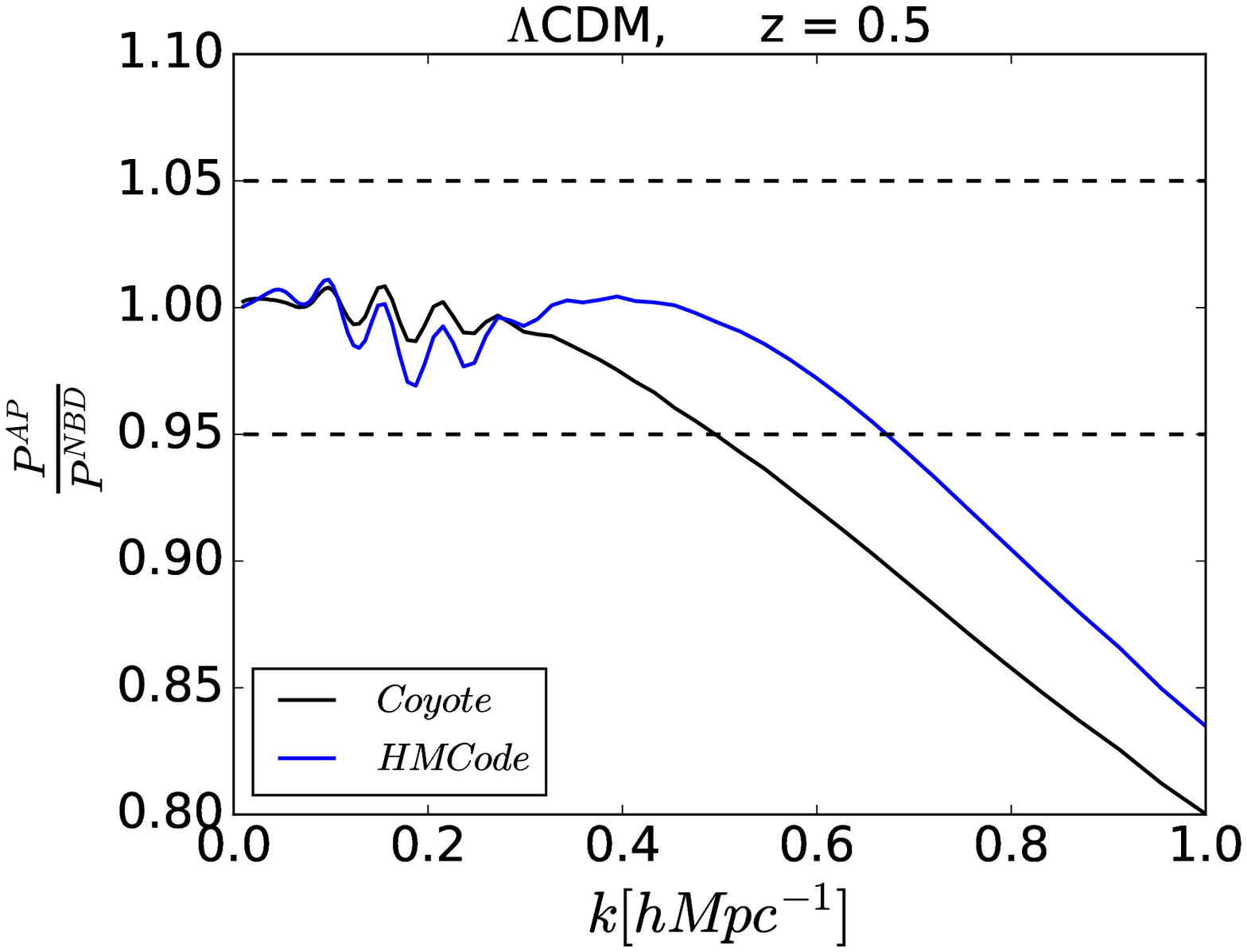,width=5.5 cm}
\epsfig{file=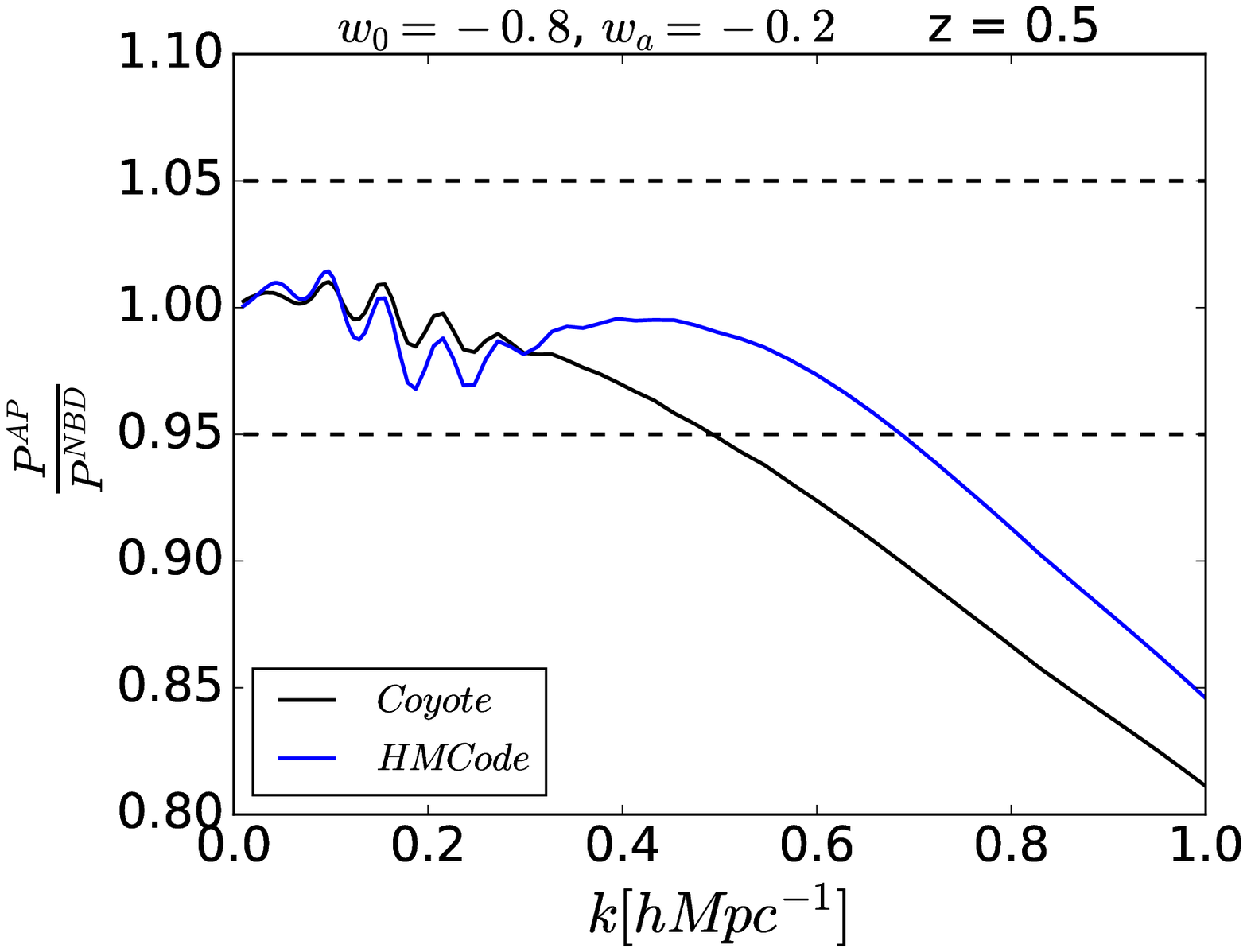,width=5.5 cm}\\
\epsfig{file=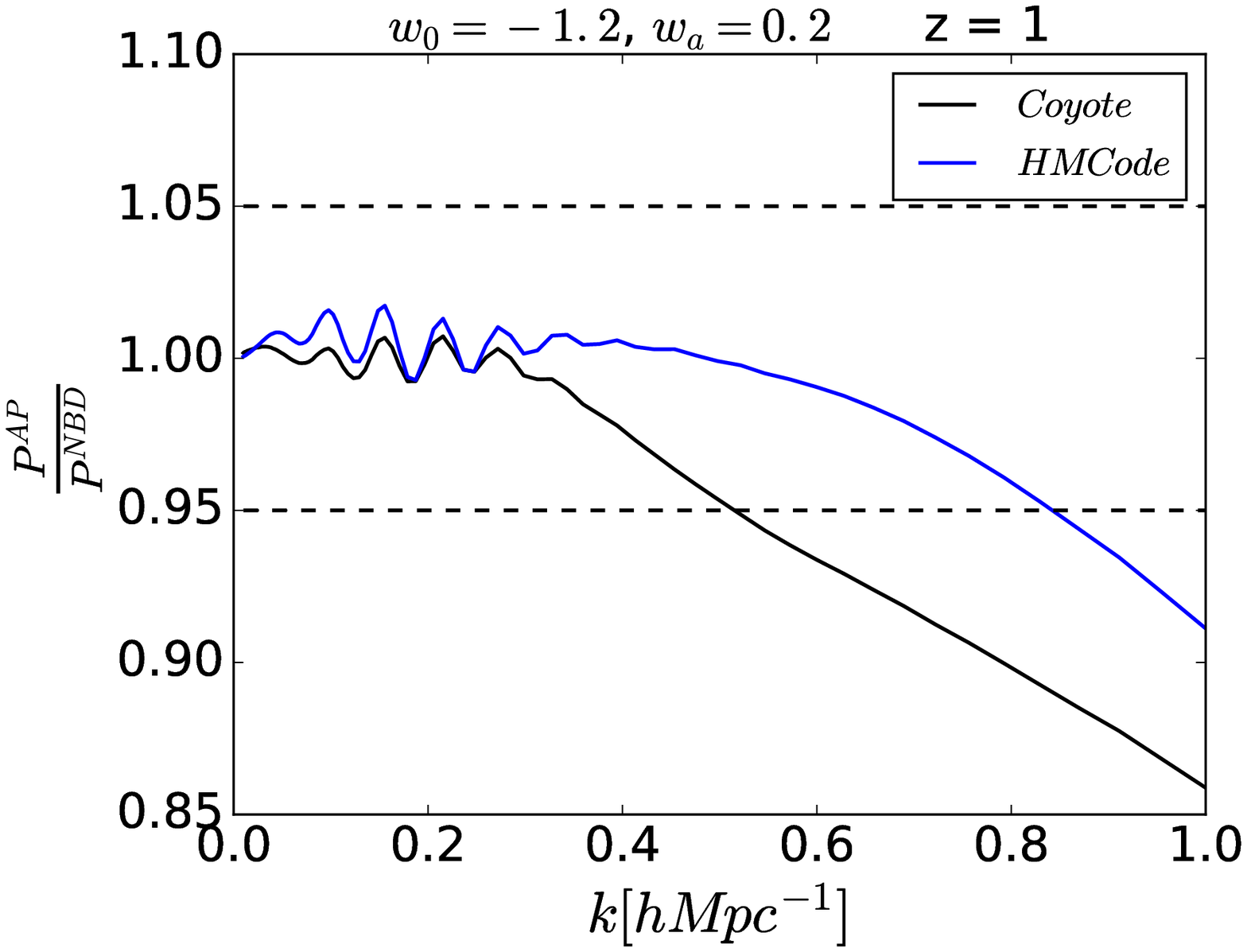,width=5.5 cm}
\epsfig{file=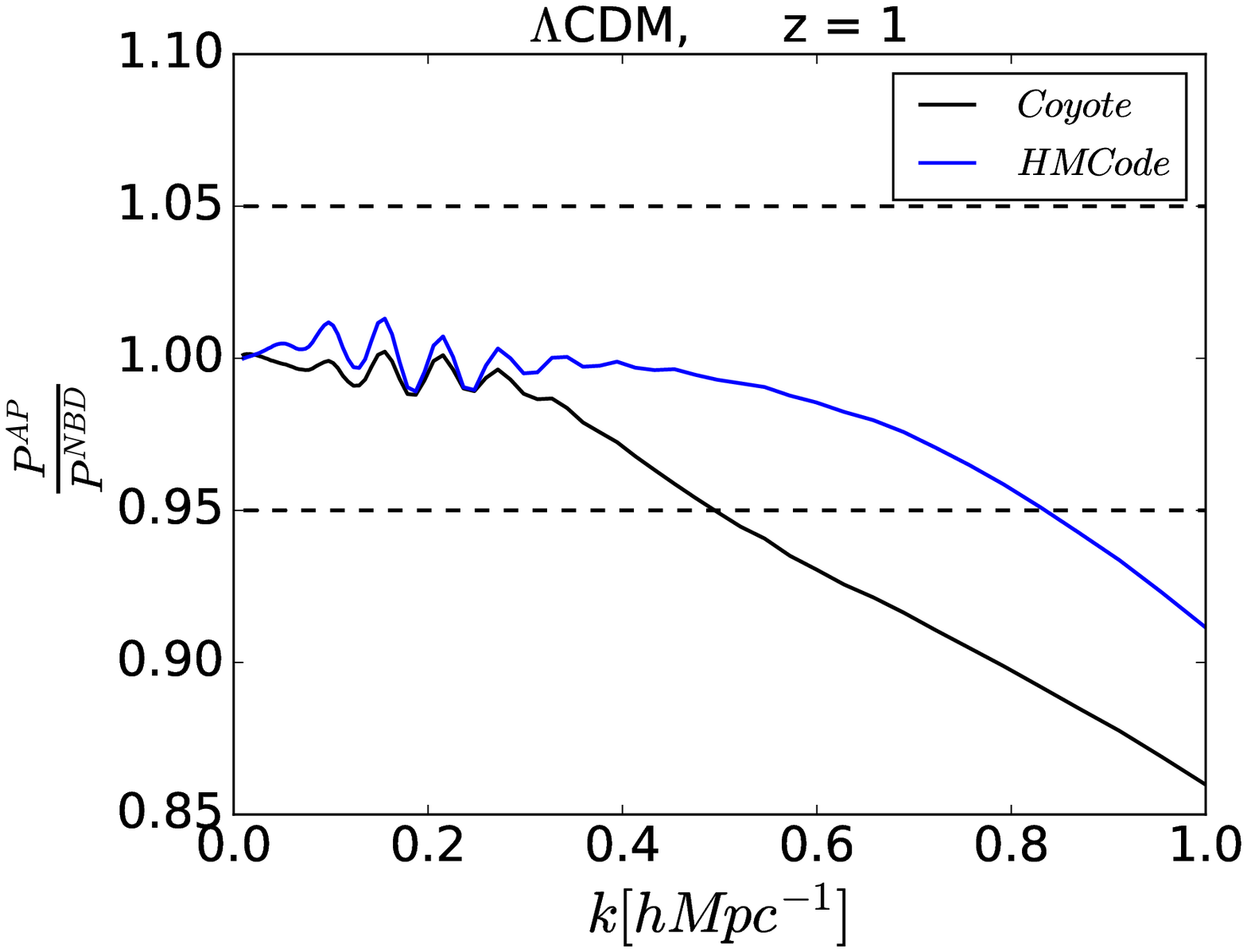,width=5.5 cm}
\epsfig{file=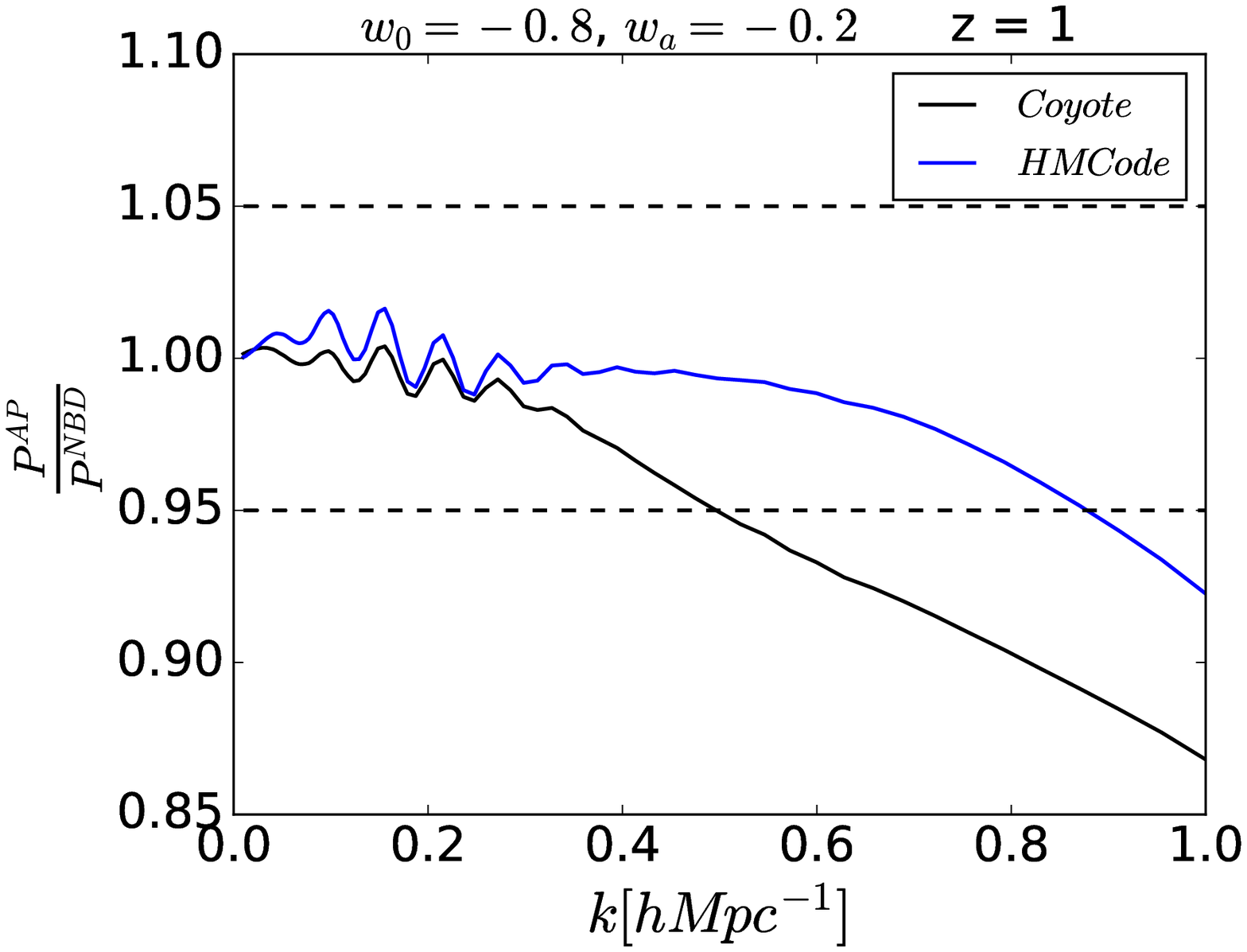,width=5.5 cm}\\
\epsfig{file=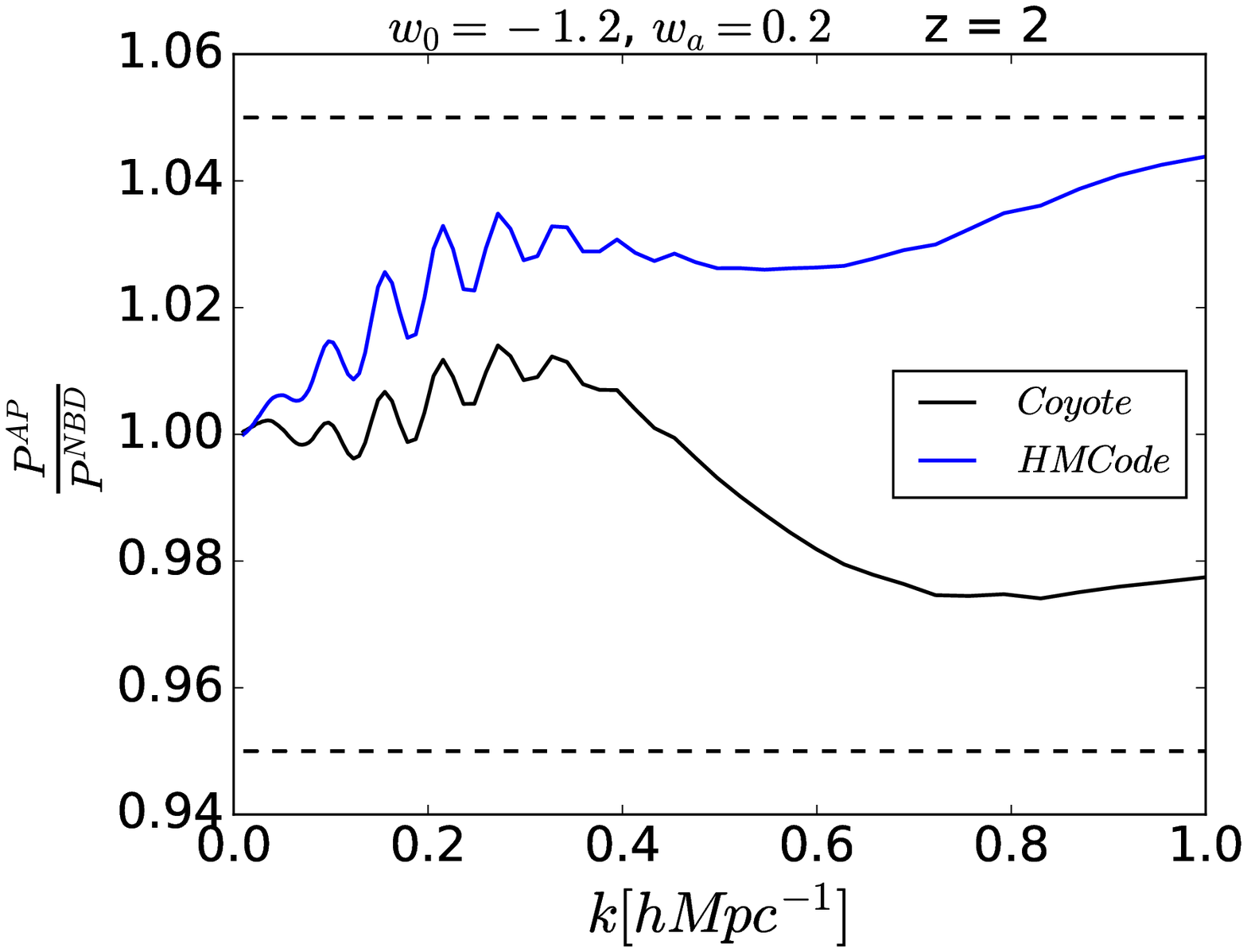,width=5.5 cm}
\epsfig{file=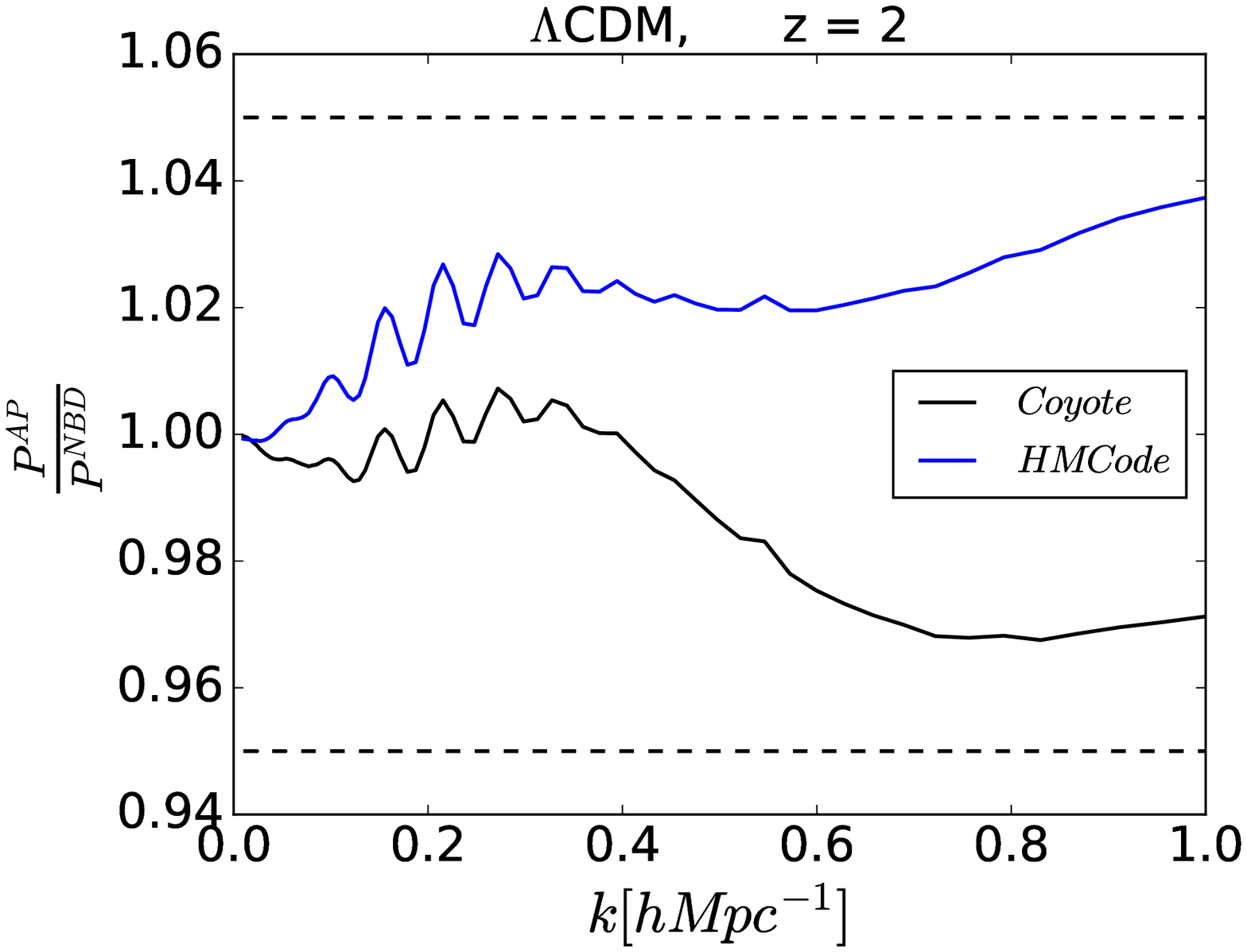,width=5.5 cm}
\epsfig{file=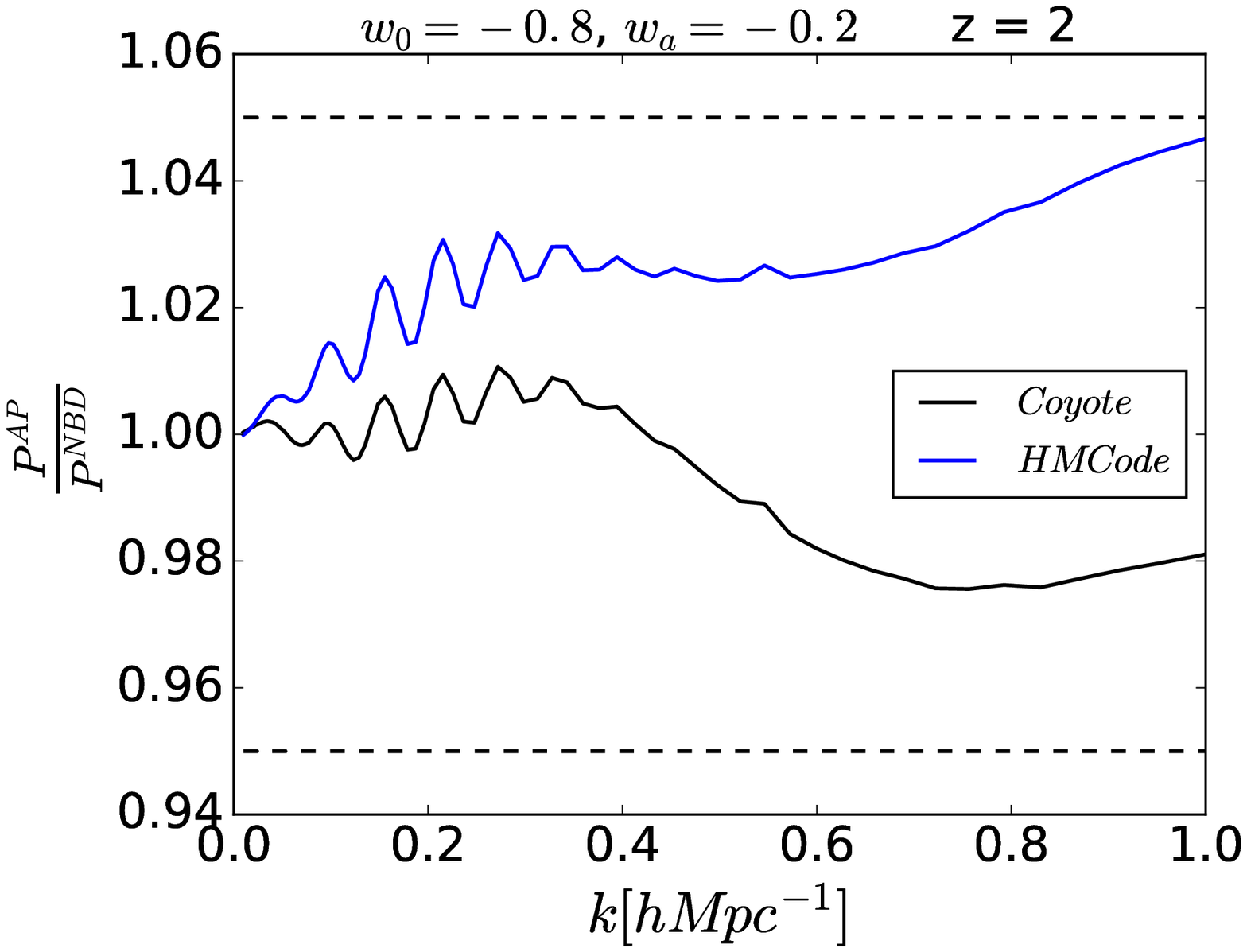,width=5.5 cm}
\end{tabular}
\caption{\label{fig:NBDcmp}Comparison between resummation scheme and two other available tools to compute nonlinear power spectrum. Unlike in other figures, here, each panel corresponds to a particular model.
}
\end{figure*}
\end{center}

\noindent
Here we have introduced a filter function $ F(k) $ which plays a role to switch off the second term of Eq. \eqref{eq:PhiTildewithFilter} for small k values as because this term contributes only for large k limit and the filter function is defined as (S. Anselmi 2012, S. Anselmi 2014)

\begin{equation}
F(k) = \frac{(k/\bar{k})^4}{1+(k/\bar{k})^4},
\label{eq:Filter}
\end{equation}

\noindent
where $ \bar{k} $ is the scale of k at which two terms of Eq. \eqref{eq:PhiTildewithFilter} are equal at present time i,e; at redshift $ z=0 $ and this corresponds to $ \bar{k} \simeq 0.2 h {Mpc}^{-1} $. The form of the function, $F(k)$ is taken in such a way that it is $0$ for $k<\bar{k}$ and when $k$ increases above $\bar{k}$ it rapidly increases towards $1$. In this way, it switches off the second term of Eq. \eqref{eq:PhiTildewithFilter} for $k<\bar{k}$. Now, how accurate is the result in Eq. \eqref{eq:PhiTildewithFilter} or in Eq. \eqref{eq:PSevolutionAppx} (accordingly) when $k$ is nearly equal to $\bar{k}$, can be seen through Fig.~\ref{fig:NBDcmp}. We shall discuss this in section 6. To mention, the errors are less than $5\%$ at lower redshifts around $\bar{k}$ and the errors decrease with increasing redshifts.
\\
Now onwards we will solve Eqs. \eqref{eq:GevolutionAppx} and \eqref{eq:PSevolutionAppx} numerically to see the behaviour of the non-linear power spectrum. Before showing numerical solution of the full non-linear power spectrum, we will shortly discuss some approximate analytical solutions of the non-linear power spectrum in the next sub-section.

\subsection{Approximate analytical solutions}
Before going to any complicacy described in the sub-section (5.2), we can get a first insight about the non-linear behavior of the clustering quintessence using the procedure described in (M. Crocce 2006c) (also see (M. Crocce 2006a, S. Anselmi 2014)) and our starting point will be Eq. \eqref{eq:doubleEvolution} which has the formal solution given by

\begin{eqnarray}
&& \psi_{a}(\vec{k},\eta) = g_{ab}(\eta,\eta_{in}) \psi_{b}^{L} (\vec{k},\eta_{in}) + \int_{\eta_{in}}^{\eta} ds \frac{e^s}{C(s)} \nonumber\\
&& g_{ab}(\eta,s) \gamma_{bcd}(\vec{k},-\vec{p},-\vec{q}) \psi_{c}(\vec{p},s) \psi_{d}(\vec{q},s),
\label{eq:PsiSol}
\end{eqnarray}

\noindent
where, we have assumed growing mode initial conditions and $ \delta_{0} $ be the initial value of the field, which satisfies $ g_{ab}(\eta,\eta_{in}) \psi_{b}^{L} (\vec{k},\eta_{in}) = \delta_{0}(\vec{k}) u_{b} $. Following (M. Crocce 2006c) $ \psi_{a}(\vec{k},\eta) $ can be expanded as

\begin{equation}
\psi_{a}(\vec{k},\eta) = \sum_{n=1}^{\infty} \psi_{a}^{(n)}(\vec{k},\eta),
\label{eq:PsiLinearcmb}
\end{equation}

\noindent
with

\begin{eqnarray}
&& \psi_{a}^{(n)}(\vec{k},\eta) = \int d^{3}q_{1} ... d^{3}q_{n} \delta_{D}(\vec{k}-\vec{q}_{1...n}) \nonumber\\
&& \mathcal{F}_{a}^{(n)}(\vec{q}_{1},...,\vec{q}_{n};\eta) \delta_{0}(\vec{q}_{1})...\delta_{0}(\vec{q}_{n}),
\label{eq:Psinthorder}
\end{eqnarray}

\noindent
where $ \vec{q}_{1...n} = \vec{q}_{1}+...+\vec{q}_{n} $. Now replacing Eqs. \eqref{eq:PsiLinearcmb} and \eqref{eq:Psinthorder} into Eq. \eqref{eq:PsiSol} we get kernel recursion relations as

\begin{eqnarray}
&& \mathcal{F}_{a}^{(n)}(\vec{q}_{1},...,\vec{q}_{n};\eta) \delta_{D}(\vec{k}-\vec{q}_{1...n}) = \Big{[} \sum_{m=1}^{n} \int_{\eta_{in}}^{\eta} ds \frac{e^s}{C(s)} \nonumber\\
&& g_{ab}(\eta,s) \gamma_{bcd}(\vec{k},-\vec{q}_{1...m},-\vec{q}_{m+1...n}) \nonumber\\
&& \mathcal{F}_{c}^{(m)}(\vec{q}_{1},...,\vec{q}_{m};s) \mathcal{F}_{d}^{(n-m)}(\vec{q}_{m+1},...,\vec{q}_{n};s) \Big{]}_{sym},
\label{eq:KernelFnthorder}
\end{eqnarray}

\noindent
where the r.h.s. of the above equation has to be symmetrized under interchange of any two wave vectors. For a $ \Lambda $CDM cosmology ($ C = 1 $) we recover the well known SPT recursion relation derived in (M. Crocce 2006c). When $ C $ is constant i.e. independent of time, we can perform the time integrals in the above equation analytically but  when $ C $ is function of time which is the case of clustering quintessence we can not analytically perform the time integrals, but if we take into account just the growing mode propagator then the propagator becomes independent of time (same happens if we take decaying mode propagator) and the time integrals become in the form as $ \int_{\eta_{in}}^{\eta} ds \frac{e^s}{C(s)} $, which can be easily integrated out using Eq. \eqref{eq:curlyI}. Following all the time integrals we can write

\begin{equation}
\mathcal{F}_{a}^{(n)}(\vec{k}_{1},...,\vec{k}_{n};\eta) \delta_{D}(\vec{k}-\vec{k}_{1...n}) \sim D_{m}^{(n-1)}(\eta).
\label{eq:KernelSolAppx}
\end{equation}

\noindent
Therefore the n-th order contribution of the total density contrast follows $ \delta_{\vec{k}}^{(n)} \sim DD_{m}^{(n-1)}  $ which leads to $ P^{(n)} \sim D^{2}D_{m}^{2(n-1)}  $, where $ P $ is defined as $ P = < \delta^{*} \hspace{0.1 cm} \delta > $.

\noindent
From the above equation we can see that in the linear regime ($ n=1 $) power spectrum is driven by total linear growth function which comes from dark matter and dark energy together and in the non-linear regime ($ n>1 $) extra effects over linearity i.e non-linearity comes from  dark matter growth function only. So, the dark energy is acting in the same way both in the linear and non-linear regimes but as we go from linear to non-linear regime extra effects come from the dark matter only.
\\
\noindent
We can also see the same result through resummation technique described in sub-section (5.2). We already have the expression of the linear power spectrum in our hand through Eq. \eqref{eq:PSlinear}. In the large k limit putting $ F(k) \rightarrow 1 $ in Eq. \eqref{eq:PhiTildewithFilter} and using Eq. \eqref{eq:GeikAppx}, the last line of Eq. \eqref{eq:PSevolutionAppx} can be analytically integrate out to become (see Eq. (B.10) in (S. Anselmi 2014))

\begin{eqnarray}
&& \int_{\eta_{in}}^{\eta} ds \Big{[} \tilde{\Phi}_{ac}(k;\eta,s)\bar{G}_{bc}(k;\eta,s) \nonumber\\
&& + \bar{G}_{ac}(k;\eta,s)\tilde{\Phi}_{cb}(k;\eta,s) \Big{]} \nonumber\\
&& = 2 u_{a} u_{b} P_{in}(k) \frac{e^{\eta}}{C(\eta)} k^2 \sigma_{v}^2 \mathcal{I}(\eta,\eta_{in}).
\label{eq:LastLinePSevlvSol}
\end{eqnarray}

\noindent
In the large k limit we can also get an analytical expression for $ H_{a}(k;\eta,\eta_{in}) $ as (see Eq. (B.3) in (S. Anselmi 2014))

\begin{equation}
H_{a}(k;\eta,\eta_{in}) = u_{a} \frac{e^{\eta}}{C(\eta)} \Big{[} - k^2 \sigma_{v}^2 \mathcal{I}(\eta,\eta_{in}) \Big{]}.
\label{eq:HSolAppx}
\end{equation}

\noindent
Using Eqs. \eqref{eq:LastLinePSevlvSol} and \eqref{eq:HSolAppx}, the differential equation \eqref{eq:PSevolutionAppx} can be solved to get power spectrum in the large k limit as (S. Anselmi 2012)

\begin{equation}
P_{ab} \rightarrow \frac{\sqrt{\pi}}{2} y (1 - \frac{1}{y \sqrt{\pi}} + \frac{1}{y^{2}}) u_{a} u_{b} P_{in},
\label{eq:PabSolAppx}
\end{equation}

\noindent
where $ y = k \sigma_{v} \mathcal{I}(\eta,\eta_{in}) $. From Eq. \eqref{eq:curlyI} we know that $ \mathcal{I} $ depends only on the matter growth function. So, it is clear that in the non-linear regime extra effects are driven by the dark matter only and the dark energy plays the role in the same way both in the linear and non-linear regimes i.e. multiplied by the square of total linear growth function which can be seen through Eq. \eqref{eq:doublet}.

\begin{center}
\begin{figure*}[!h]
\begin{tabular}{c@{\quad}c}
\epsfig{file=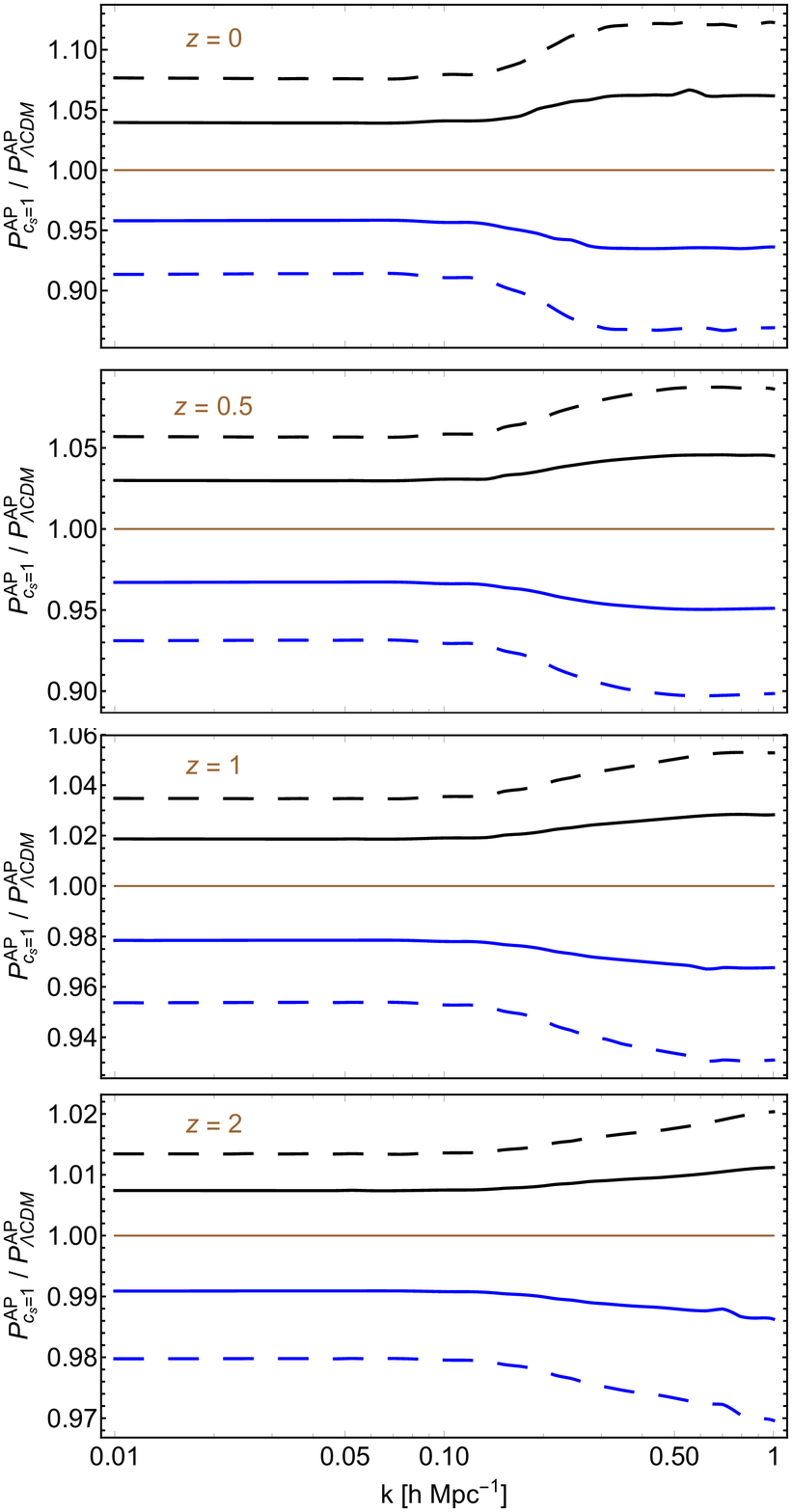,width=7.5 cm}
\epsfig{file=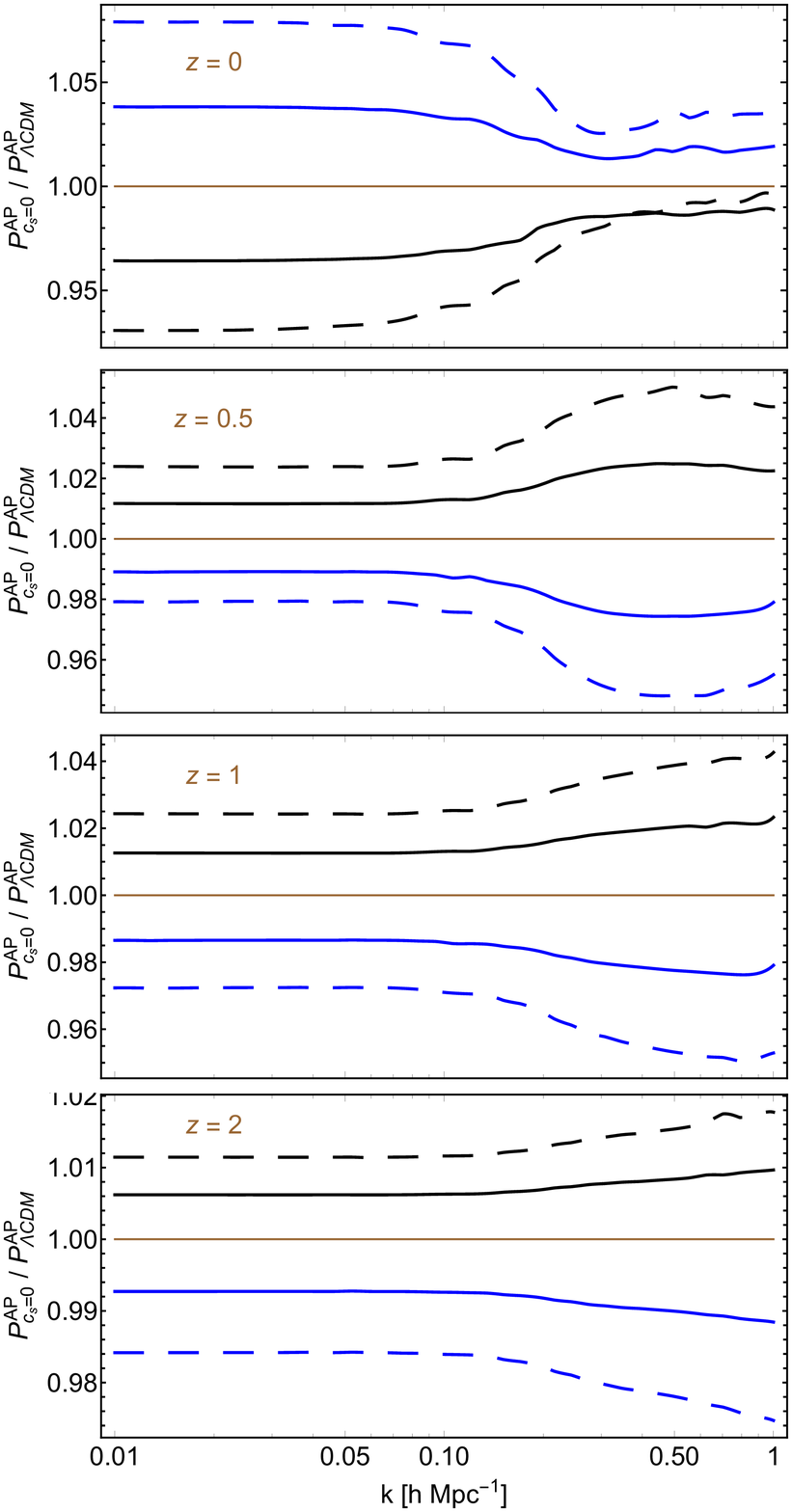,width=7.5 cm}
\end{tabular}
\caption{\label{fig:Pcs01byPSL} PS compared to $ \Lambda $CDM as function of $ k $. 
Left Panel: ($ c_{s} = 1 $). 
Right Panel: ($ c_{s} = 0 $).
The color code is same as in Fig.~\ref{fig:wofz}.
}
\end{figure*}
\end{center}

\begin{figure}
\begin{center}
\includegraphics[scale=0.5]{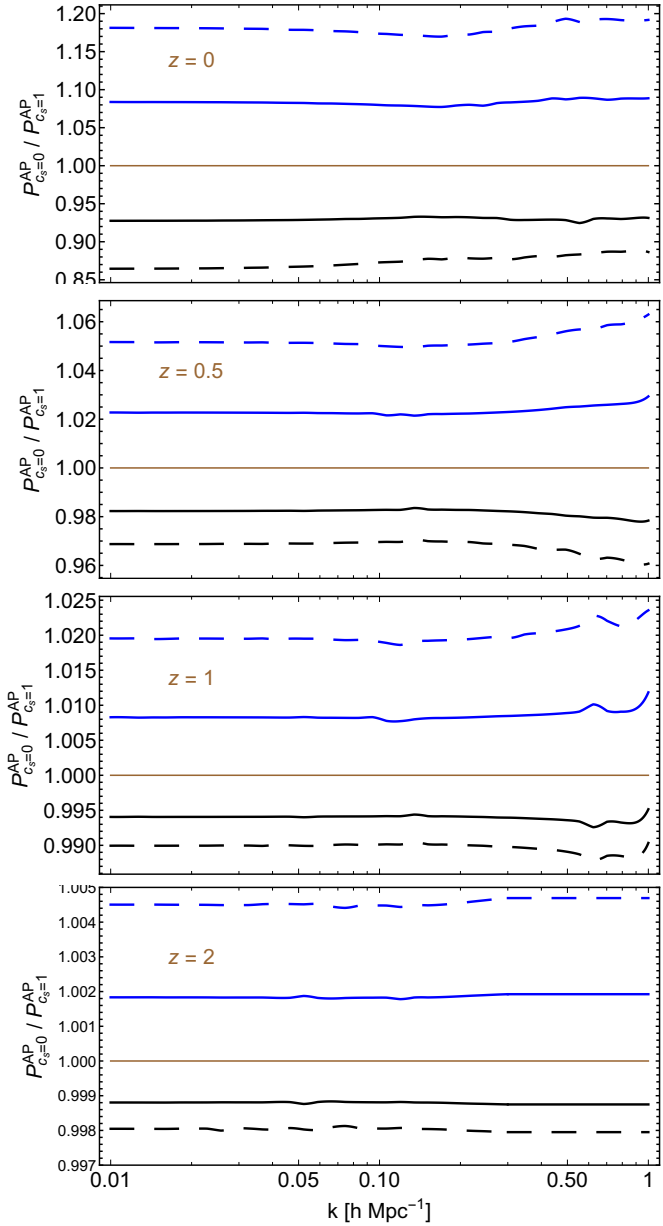}
\end{center}
\caption{\label{fig:Pcs0byPcs1} $ PS(k) $ vs. $ k $ graphs: (comparison between  $ c_{s} = 0 $ and $ c_{s} = 1 $ cases). The color code is same as in Fig.~\ref{fig:wofz}.}
\end{figure}

\section{Results}

We now solve Eq. \eqref{eq:PSevolutionAppx} numerically to get the solution of the non-linear total density power spectrum both for smooth and clustering quintessence and present the results in Figs.~\ref{fig:Pcs01byPSL} and ~\ref{fig:Pcs0byPcs1}. To solve Eq. \eqref{eq:PSevolutionAppx} we have taken same value of the amplitude of the initial power spectrum for all the cases and the initial conditions are taken at redshift $ z_{in} =1000 $. The initial power spectrum is computed by evolving back the $ \Lambda $CDM linear power spectrum from $ z=0 $ (using CAMB (A. Lewis 2000)) to $ z_{in} $ using linear theory of cosmological perturbations in the Newtonian approximation where we have taken $ \Omega_{m}^{(0)} h^{2} = 0.13263 $, $ \Omega_{b}^{(0)} h^{2} = 0.02273 $, $ h = 0.72 $, $ \sigma_{8} = 0.79 $ and $ n_{s} = 0.963 $. Here, we have considered a transfer function which is default in the CAMB i.e. the halo fit model for the transfer function.

In Fig.~\ref{fig:Pcs01byPSL}, we have plotted power spectrum compared to $ \Lambda $CDM one both for smooth and clustering dark energy. In Fig.~\ref{fig:Pcs0byPcs1}, we have plotted power spectrum for $ c_{s} = 0 $ compared to $ c_{s} = 1 $ as function of $ k $ for the same model respectively. In Fig.~\ref{fig:Pcs01byPSL}, left and right panels are for $ c_{s} = 1 $ and $ c_{s} = 0 $ respectively. Both in Figs.~\ref{fig:Pcs01byPSL} and~\ref{fig:Pcs0byPcs1}, the color code is same as in Fig.~\ref{fig:wofz} for all the models respectively. Hereafter in all the figures we define $ P $ as total power spectrum which is given by $ P = < \delta^{*} \hspace{0.1 cm} \delta > = e^{2 \eta} < \psi_{1}^{*} \hspace{0.1 cm} \psi_{1} > = D^2 P_{11} $. In all the figures, superscript 'AP' corresponds to the resummation results, where we numerically solve Eqs. \eqref{eq:PSevolutionAppx} to get the nonlinear power spectrum.

First of all in the linear regime ($ k < 0.2 h Mpc^{-1} $) power spectrum are independent of $ k $ which is exactly expected to be in linear perturbation theory and the amplitudes of the total power spectrums are completely consistent with results correspond to the Fig.~\ref{fig:DLcmp}. But in the non-linear regime the power spectrums depend on $ k $ drastically. At low redshift we can neglect last two terms of Eq. \eqref{eq:PabSolAppx} as $ y >> 1 $ which corresponds to $ P_{11} \propto y $ and we can also approximate $ \mathcal{I}(\eta,\eta_{in}) \simeq \frac{D_{m}(\eta)}{D_{in}} $ which gives $ P_{11} \propto D_{m} $. So, we get for small k limit $ P \propto D^{2} $ and for large k limit $ P \propto D^{2} D_{m} $.

Now, from the above discussions, about the left panel Fig.~\ref{fig:Pcs01byPSL}, we can say that $ \dfrac{P_{cs=1}}{P_{\Lambda CDM}} \propto \left( \dfrac{D_{cs=1}}{D_{\Lambda CDM}} \right)^{2} $ on large scales and $ \dfrac{P_{cs=1}}{P_{\Lambda CDM}} \propto \left( \dfrac{D_{cs=1}}{D_{\Lambda CDM}} \right)^{2} \left( \dfrac{D_{m,cs=1}}{D_{m,\Lambda CDM}} \right) = \left( \dfrac{D_{cs=1}}{D_{\Lambda CDM}} \right)^{3} $ on small scales. So, using this logic, we can see that top-left panel of Fig.~\ref{fig:DLcmp} and left panel of Fig.~\ref{fig:Pcs01byPSL} are almost consistent. We can use the same logic from the previous discussions and one can check that right panel of Fig.~\ref{fig:Pcs01byPSL} and Fig.~\ref{fig:Pcs0byPcs1} are also consistent with Fig.~\ref{fig:DLcmp} accordingly. So, the physics is the same as in the discussions of the Fig.~\ref{fig:DLcmp}.

From top-left and top-right panels of the Fig.~\ref{fig:Pcs01byPSL}, it seems that there is an ambiguity between phantom smooth and non-phantom clustering (and vice-versa) in the matter power spectrum at lower redshifts. However, the behavior is not exactly same for all scales. One can see the power spectrum increases (decreases) with $k$ for phantom smooth (non-phantom smooth) whereas it decreases (increases) for non-phantom clustering (phantom clustering). At higher redshifts, there is no ambiguity. Although there is ambiguity between two cases at lower redshifts, the background dynamics are completely different. So, if we combine background and perturbations observables together, there will be no ambiguity at all.

In Fig.~\ref{fig:NBDcmp}, we compare our resummation results for the nonlinear power spectrum for the smooth case with other two available tools. One is HMCode (A. Mead 2016) and another is the Coyote Emulator (L. Casarini 2009, Casarini 2016, Katrin Heitmann 2014, Katrin Heitmann 2010, Katrin Heitmann 2009, Earl Lawrence 2010). We can see that the resummation results are at few percentage accuracies compared to the HMCode or Coyote Emulator up to $ k < 0.6 h Mpc^{-1} $ at smaller redshifts. At higher redshifts, the accuracies increase. Note that we have compared the results taking three models among the 5 models considered in this paper. The results are similar for the other two models which we have not included because the conclusions will be the same after including these two models. The superscript 'NBD' refers to either the HMCode (Blue lines) or the Coyote Emulator (Black lines) results respectively.

Now we want to investigate the approximate relation of the nonlinear clustering between smooth and clustering dark energy scenarios. Ref. (S. Anselmi 2011a) found that the absolute value of deviations of the nonlinear total power spectrums from linear one in both the smooth and clustering dark energy scenarios are comparable, at better than $ 1 \% $ in the BAO region. In our case, we can also guess nearly same result from Figs.~\ref{fig:Pcs01byPSL} and \ref{fig:Pcs0byPcs1} up to the BAO region. This corresponds to the approximation

\begin{equation}
P_{c_{s} = 0}(k;\eta) - P_{c_{s} = 0}^{L}(k;\eta) \simeq P_{c_{s} = 1}(k;\eta) - P_{c_{s} = 1}^{L}(k;\eta).
\label{eq:PAppx1}
\end{equation}

We can also get another approximation relation of the nonlinear total power spectrums between smooth and clustering dark energy scenarios using the result of the sub-section (5.3). First of all the linear power spectrum is proportional to the square of the linear growth function. But in the nonlinear regime power spectrum associates to some extra factor driven by linear matter growth function which can be seen through Eq. \eqref{eq:KernelSolAppx} or Eq. \eqref{eq:PabSolAppx}. Using these two limits we can write an expression for power spectrum in the intermediate k regime introducing two filters as $ F(k) $ in the large k term and $ 1-F(k) $ for small k term which corresponds to the approximation

\begin{equation}
P_{c_{s}=0} \simeq \Big{[}\frac{D_{c_{s}=0}}{D_{c_{s}=1}}\Big{]}^{2} P_{c_{s}=1}  \Big{[} (1-F(k))+ F(k) \Big{(}\frac{D_{m},_{c_{s}=0}}{D_{m},_{c_{s}=1}} \Big{)} \Big{]}.
\label{eq:PAppx2}
\end{equation}

\begin{center}
\begin{figure*}[!h]
\begin{tabular}{c@{\quad}c}
\epsfig{file=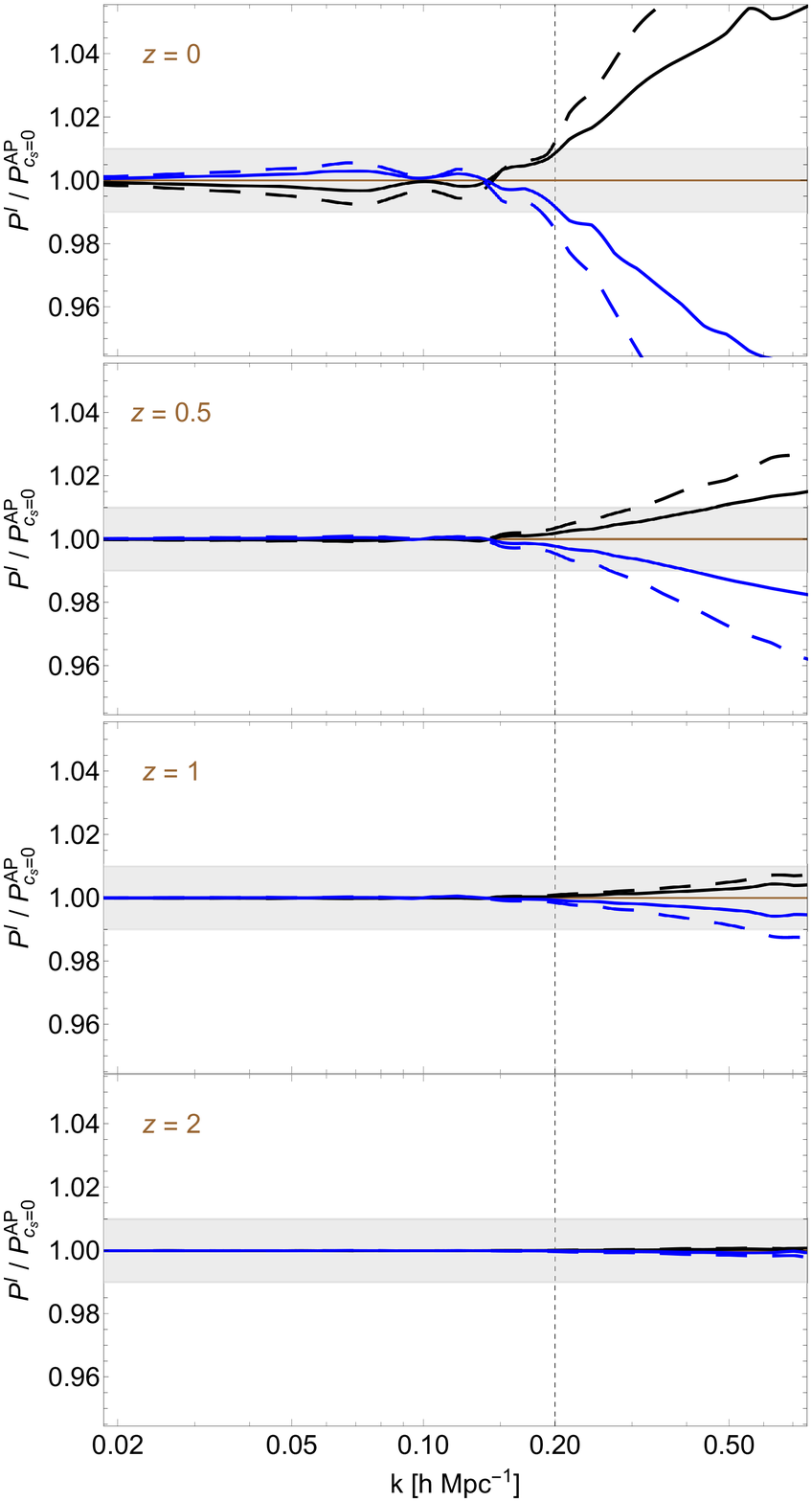,width=7.5 cm}
\epsfig{file=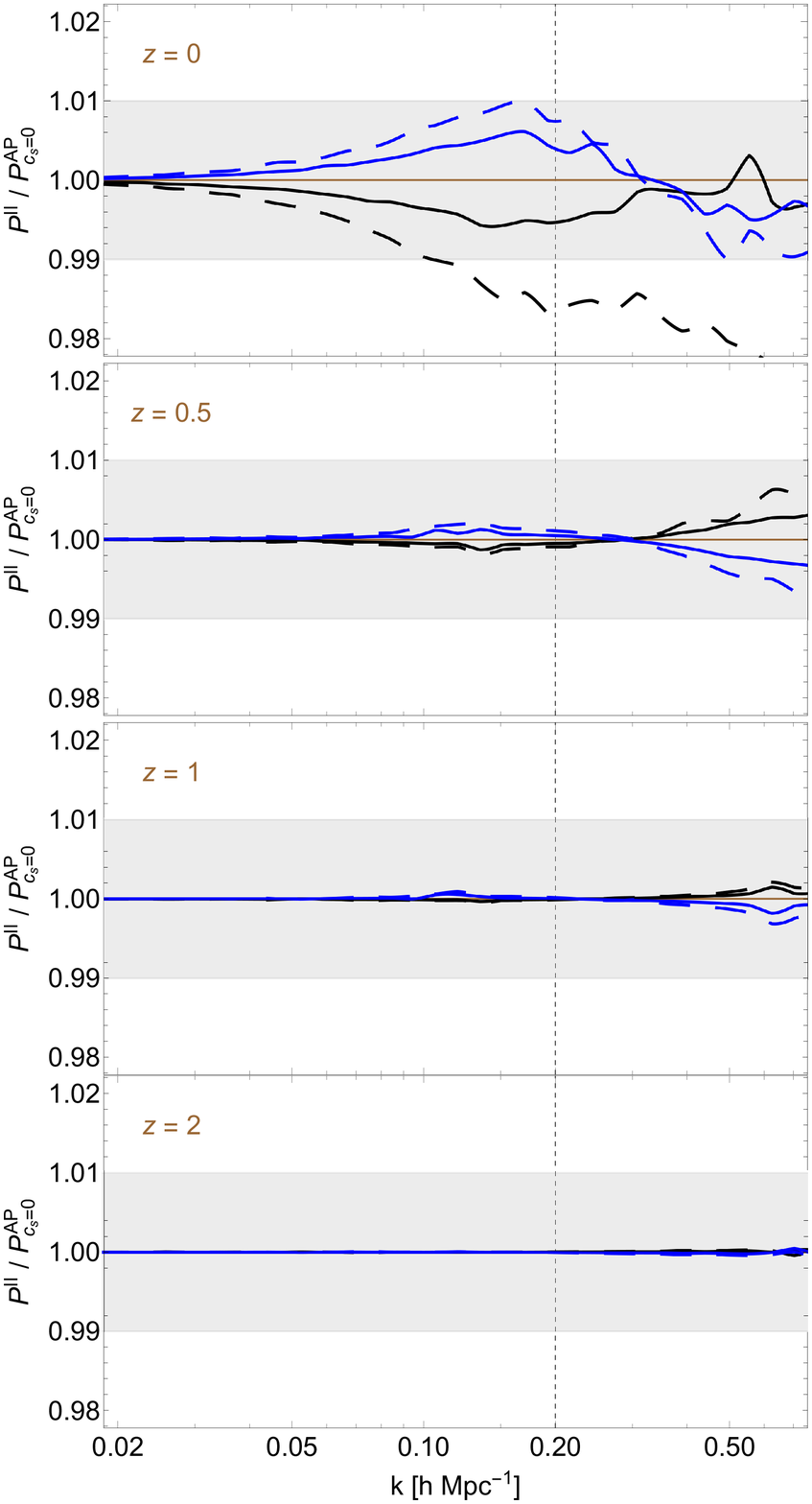,width=7.5 cm}
\end{tabular}
\caption{\label{fig:PSappx} Compared PS as function of $ k $. 
Left Panel: (Using Eq. \eqref{eq:PAppx1} and denoted by $ "I" $). 
Right Panel: (Using Eq. \eqref{eq:PAppx2} and denoted by $ "II" $).
The color code is same as in Fig.~\ref{fig:wofz}.
}
\end{figure*}
\end{center}

\noindent
In Fig.~\ref{fig:PSappx}, we have shown how accurate the above two approximations holds. In the left Panel we have used Eq. \eqref{eq:PAppx1} and denoted it by $ "I" $ and in the right Panel we have used Eq. \eqref{eq:PAppx2} and denoted it by $ "II" $. In both the panels, the color code is same as in Fig.~\ref{fig:wofz} for all the models respectively. To plot Fig.~\ref{fig:PSappx}, we take resummation results for $ c_{s} =1 $ case and put these values into Eqs. \eqref{eq:PAppx1} and \eqref{eq:PAppx2} and compute the power spectrum for $ c_{s} =0 $ case according to these equations (which we denote by superscript 'I' and 'II' respectively) and compare these results to the resummation result for $ c_{s} =0 $ case (which we denote by superscript 'AP'). So, approximation \eqref{eq:PAppx1} is good upto BAO region but the approximation \eqref{eq:PAppx2} is better upto an accuracy $ 1 \% $ both in the linear and non-linear regimes and both these approximations are more accurate as we go to high redshifts. We believe that approximate Eqs. \eqref{eq:PAppx1} and \eqref{eq:PAppx2} can be applied to any other tools at few percentage accuracies i.e. we can compute nonlinear power spectrum for the smooth case from the any other standard tools (instead of the resummation scheme) and using Eqs. \eqref{eq:PAppx1} and \eqref{eq:PAppx2} we can find nonlinear power spectrum for the clustering dark energy models. Especially the approximation \eqref{eq:PAppx2} is more useful.

As because approximation \eqref{eq:PAppx2} is accurate enough to an extent of $ 1\% $ level we can assume similar approximation holds good to any value of k for dark energy models with a generic speed of sound $ c_{s} $ as 

\begin{equation}
P_{c_{s}} \simeq \Big{[}\frac{D_{c_{s}}}{D_{c_{s}=1}}\Big{]}^{2} P_{c_{s}=1}  \Big{[} (1-F(k))+ F(k) \Big{(}\frac{D_{m},_{c_{s}}}{D_{m},_{c_{s}=1}} \Big{)} \Big{]}.
\label{eq:PAppx3}
\end{equation}

\noindent
So, although it is much more complicated to find non-linear solutions for dark energy models with a generic speed of sound $ c_{s} $ we can get an approximate solution to these models to an extent using Eq. \eqref{eq:PAppx3} which can be tested in future work.

\section{Conclusion}

First of all, we compute the nonlinear power spectrum by solving Eq. \eqref{eq:PSevolutionAppx} numerically in the mildly nonlinear regime both for clustering and smooth dark energy with evolving equation of state. There are no available N-body simulations (popular) for clustering dark energy models. However, we compare our results of the smooth dark energy with HMCode (A. Mead 2016) and Coyote Emulator (L. Casarini 2009, Casarini 2016, Katrin Heitmann 2014, Katrin Heitmann 2010, Katrin Heitmann 2009, Earl Lawrence 2010) results. The results are at few percentage accuracies up to $ k < 0.6 h Mpc^{-1} $ at smaller redshifts. At higher redshifts, the accuracies increase.

We have shown that the presence of dark energy affects the clustering in the same way both on small and large scales. This conclusion is valid both for smooth and clustering quintessence. The small-scale clustering differs from the large-scale clustering due to the clustering in the matter part only.

Since the dark energy affects the clustering in the same way on all the scales, we were able to present some approximate simple relations of nonlinear power spectrum between smooth and clustering quintessence as in Eq. \eqref{eq:PAppx1} or in Eq. \eqref{eq:PAppx2}. The approximate relation \eqref{eq:PAppx1} is at few percentage accuracies up to BAO scale whereas the relation \eqref{eq:PAppx2} is more accurate up to larger values of $k$. The accuracies increase with increasing redshifts for both the approximations.

Finally, we present another approximate semi-analytical expression for nonlinear power spectrum for an arbitrary sound speed of dark energy $0< c_{s} <1$ in Eq. \eqref{eq:PAppx3}. The results of the effects of dark energy with $0< c_{s} <1$ need to be verified in the future works.

In summary, our numerical results from Eq. \eqref{eq:PSevolutionAppx} or the semi-analytical results from the approximate Eqs. \eqref{eq:PAppx1} and \eqref{eq:PAppx2} can give an idea about the nonlinear power spectrum at the few percentage accuracies up to $k \lesssim 0.6 h Mpc^{-1}$ at smaller redshifts both for smooth and clustering quintessence. The accuracies increase with increasing redshifts. We also present another approximate relation in Eq. \eqref{eq:PAppx3} to compute nonlinear power spectrum for the dark energy with generic speed of sound where $0< c_{s} <1$.

\section*{Acknowledgements}

The author would like to acknowledge Council of Scientific $\&$ Industrial Research (CSIR), Govt. of India for financial support through Senior Research Fellowship (SRF) scheme No:09/466(0157)/2012-EMR-I. The author would like to thank Stefano Anselmi, Anjan Ananda Sen and Diana Lo'pez Nacir for the exclusive discussions. The author would also like to thank Pier-Stefano Corasaniti and Massimo Pietroni.
\vspace{-1em}



\end{document}